\documentclass[11pt,a4paper]{article}

\usepackage{jheparxiv}
\usepackage{basket}
\usepackage{mathShrtcts}
\usepackage{slashed}

\title{A Gluing Operator for the Ambitwistor String}

\author{Kai A. Roehrig and David Skinner}
\affiliation{Department of Applied Mathematics \& Theoretical Physics \\
        University of Cambridge \\
        Wilberforce Road \\
        Cambridge CB3 0WA, United Kingdom}
\emailAdd{kafr2@cam.ac.uk, d.b.skinner@damtp.cam.ac.uk}

\date{\today}

\abstract{We present a new operator in the ambitwistor string which glues together correlators with fewer points or of lower genus.  It underpins the recursive construction of tree-level CHY scattering amplitudes by Dolan \& Goddard, as well as the computation of loop integrands on a Riemann sphere by Geyer et al. The gluing operator is a tractable object due to the finiteness of the spectrum. In particular, we demonstrate how it gives rise to the complete one-loop integrand in SYM and SUGRA. The operator is conjectured to be the path integral incarnation of the ambitwistor string propagator, and to coincide with the field theory limit of the standard string theory propagator.}

 \subheader{\hfill \texttt{DAMTP-2017-NN}}

\begin{document}

\maketitle


 \section{Introduction}
 
Ambitwistor strings are chiral, diffeomorphism invariant theories defined on a Riemann surface~\cite{Mason:2013sva,Berkovits:2013xba,Adamo:2013tsa,Ohmori:2015sha,Casali:2015vta}. Remarkably, their spectrum corresponds to that of a local {\it field} theory in space--time -- in the best understood case, that of $d=10$ Type II A/B supergravity~\cite{Mason:2013sva}. Worldsheet correlation functions in the ambitwistor string localise on solutions of the scattering equations~\cite{Fairlie:1972zz,Gross:1987kza,Gross:1987ar,Cachazo:2013hca} and, at genus zero, provide a theoretical framework in which to understand the remarkable CHY representation~\cite{Cachazo:2013hca,Cachazo:2013iea} of tree--level scattering amplitudes. Correspondingly, if one places ambitwistor strings on a non-linear curved background~\cite{Adamo:2014wea}, the theory is consistent iff the background obeys the supergravity equations of motion, with no $\alpha'$ corrections.

One can also consider ambitwistor strings at higher genus~\cite{Adamo:2013tsa,Adamo:2015hoa}. Around flat space--time, correlation functions of the worldsheet CFT again localize on solutions to higher genus scattering equations~\cite{Adamo:2013tsa,Geyer:2015bja} so that the integral over the moduli space of higher genus curves again amounts to summing the correlator over solutions to these equations, which now fix the worldsheet complex structure in addition to the location of the vertex operators.  After summing over all solutions to these equations (and, in the case of RNS-type models, summing over worldsheet spin structures) the integrand is again a rational function of the external data -- as expected for the integrand of a field theory. This rational function has been explicitly computed at four points, where for $g=1,2$ it coincides with the 1-loop~\cite{Geyer:2015bja,Geyer:2015jch} and 2-loop~\cite{Adamo:2015hoa} supergravity integrands, respectively. At higher points it has been shown that the $g=1$ ambitwistor string has the correct factorization properties~\cite{Adamo:2013tsa,Adamo:2015hoa} and correct behaviour as one of the gravitons becomes soft~\cite{Casali:2014hfa}, as expected for supergravity.  The loop integrals themselves arise from the zero-modes of worldsheet field $P_\mu(z)$. Note that since the moduli space of the ambitwistor string is non-compact, unlike that of the standard string, UV divergences can arise.

Since they arise from correlation functions on a curve of $g>0$, {\it before} summing over solutions to the scattering equations, the ambitwistor correlation functions are naturally written in terms of Riemann theta functions. One does not expect to find such theta functions appearing in a {\it field theory} amplitude, suggesting that there should be a simpler way of rewriting the ambitwistor correlation function. This was found by Geyer {\it et al.} in~\cite{Geyer:2015bja,Geyer:2015jch}, following earlier work of Dolan \& Goddard~\cite{Dolan:2013isa} related to BCFW recursion of the tree--level CHY formulae. In~\cite{Geyer:2015bja,Geyer:2015jch} it was shown that, instead of localising the ambitwistor string to solutions of the higher genus scattering equations, one could localise to the boundary of the $g=1$ moduli space corresponding to a non-separating degeneration. This is achieved using the global residue theorem in $\overline{\mathcal{M}}_{1,n}$ to take the contour to surround this boundary divisor -- where the ambitwistor string integrand again has a simple pole -- instead of one of the poles of the scattering equations. The advantage of this approach is that, on this non--separating boundary divisor, the integrand can be expressed in terms of functions on the nodal Riemann sphere, much more closely in line with what one expects from a field theory Feynman graph.

\medskip

In this paper, we propose that the loop integrands can also be computed directly from a correlation function in the $g=0$  ambitwistor string on nodal Riemann surfaces. As well as vertex operators representing the external states, the correlation function also involves a new operator $\Delta(z,w)$ that we call the {\it gluing operator}.  This gluing operator plays the role of the propagator in the target space field theory. It is surprising that an inherently off-shell object such as a field theory propagator can be represented by a BRST invariant insertions. Indeed, as discussed above, local operators in the BRST cohomology of the ambitwistor string represent {\it on-shell} states of 10d SUGRA, so since the gluing operator represents an off-shell propagator, it cannot be an element of the BRST cohomology of local operators. It seems natural to give up the condition of locality (rather than BRST invariance), and indeed the gluing operator is non-local, while it retains full BRST invariance. 

Although $\Delta(z,w)$ is genuinely non-local on $\Sigma$ (rather than just bi-local), the two points $z$ and $w$ play a special role. These two points are each associated with the insertion of a set of local operators that now correspond to `off-shell states'. The role of these operators can be understood as follows. The relation to the target space propagator dictates the role of the gluing operator in a factorization limit of the original amplitude. More precisely, unitarity demands that, whenever the corresponding target space propagator goes on-shell, the gluing operator has a simple pole with residue given by the insertion of a complete set of states in the Hilbert space. By the state-operator correspondence, this can be implemented by a sum over a complete set of local vertex operators in the BRST cohomology, representing the on-shell particle flowing out of one node and into the other. Away from the factorization channel, we must extend the local operators off-shell, and by themselves they cannot be BRST invariant. The failure of these off-shell insertions to be BRST invariant is fully compensated by the remaining, non-local pieces of $\Delta(z,w)$.

These two special points may be inserted on different curve components, corresponding to divisors in the boundary of the moduli space describing separating degenerations, or both on the same curve component for a non-separating degeneration. The insertion of $\Delta(z,w)$ operator can thus intuitively be thought of as identifying $z$ and $w$, thus changing the topology of the worldsheet, as well as transporting the CFT data from one node to the other.

 \begin{figure}[t]
 \centering
 \includegraphics[scale=0.6]{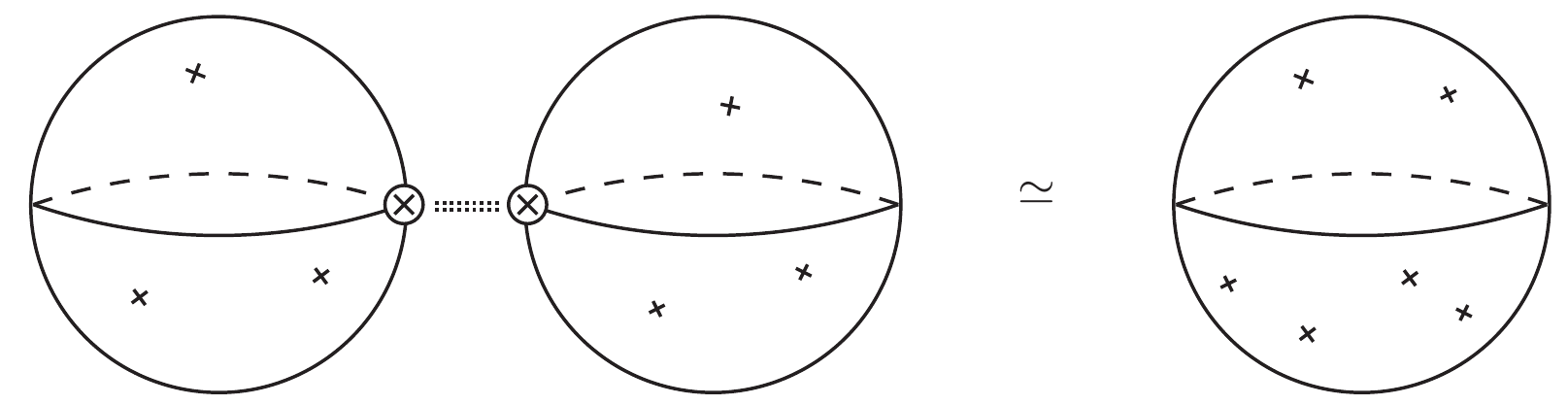}
 \caption{Gluing two genus zero correlators, forming a new sphere.}
 \label{fig:Gluing1}
 \end{figure}
 
 In particular, we claim that the full, $n$-particle tree amplitude can be computed by using $\Delta(z,z')$ to glue together two sub-amplitudes, each with one leg off-shell as
 \begin{equation}\label{eqn:GeneralFormTree}
 \int _{\overline{\mathcal{M}}_{0,n}} \!\!\!\!\left\langle    \mathcal{O}_1(z_1) \cdots  \mathcal{O}_n(z_n)   \right\rangle _\Sigma 
 = \hspace{-0.6cm}\sum_{\substack{\text{channels} \\ L \, \cup \, R = \{1, \cdots ,n \}}}  \!\!\!\!\!  \int _{\partial\overline{ \mathcal{M}}_{0,n}} 
 \!\!\! \left\langle    \mathcal{O}^{(s)}(z_\ast)  \, \prod_{i\in L}  \mathcal{O}_i(z_i)   \right\rangle _{\!\!\Sigma_L} 
  \!\!\!\!  \left\langle       \mathcal{O}_{(s)}(z'_\ast) \,\prod_{j\in R}  \mathcal{O}_j(z'_j)   \right\rangle _{\!\!\Sigma_R} \! \! ,
 \end{equation}
where $\Delta(z,z')\sim\mathcal{O}^{(s)}(z) \mathcal{O}_{(s)}(z')$ is a schematic representation of the gluing operator, whose detailed form will be given below. This operator is inserted at the nodes (with coordinates $z_\ast$ and $z'_\ast$ on $\Sigma_{L/R}$, respectively) and here has one `leg' on the left curve component and one on the right, as in~\cref{fig:Gluing1}. The correlation functions on the {\it rhs} each correspond to tree amplitudes extended to allow the leg associated to $z_\ast / z'_\ast$ to go off-shell\footnote{Off-shell continuations of CHY amplitudes have previously been considered e.g. in \cite{Dolan:2013isa,Lam:2015mgu,Lam:2015sqb}.}. The sum over channels in~\eqref{eqn:GeneralFormTree} will be explained in more detail below, but it is essentially a sum over all the different boundary divisors in $\overline{\mathcal{M}}_{0,n}$ where the original integrand had a single pole.

\begin{figure}[t]
 \centering
 \includegraphics[scale=0.5]{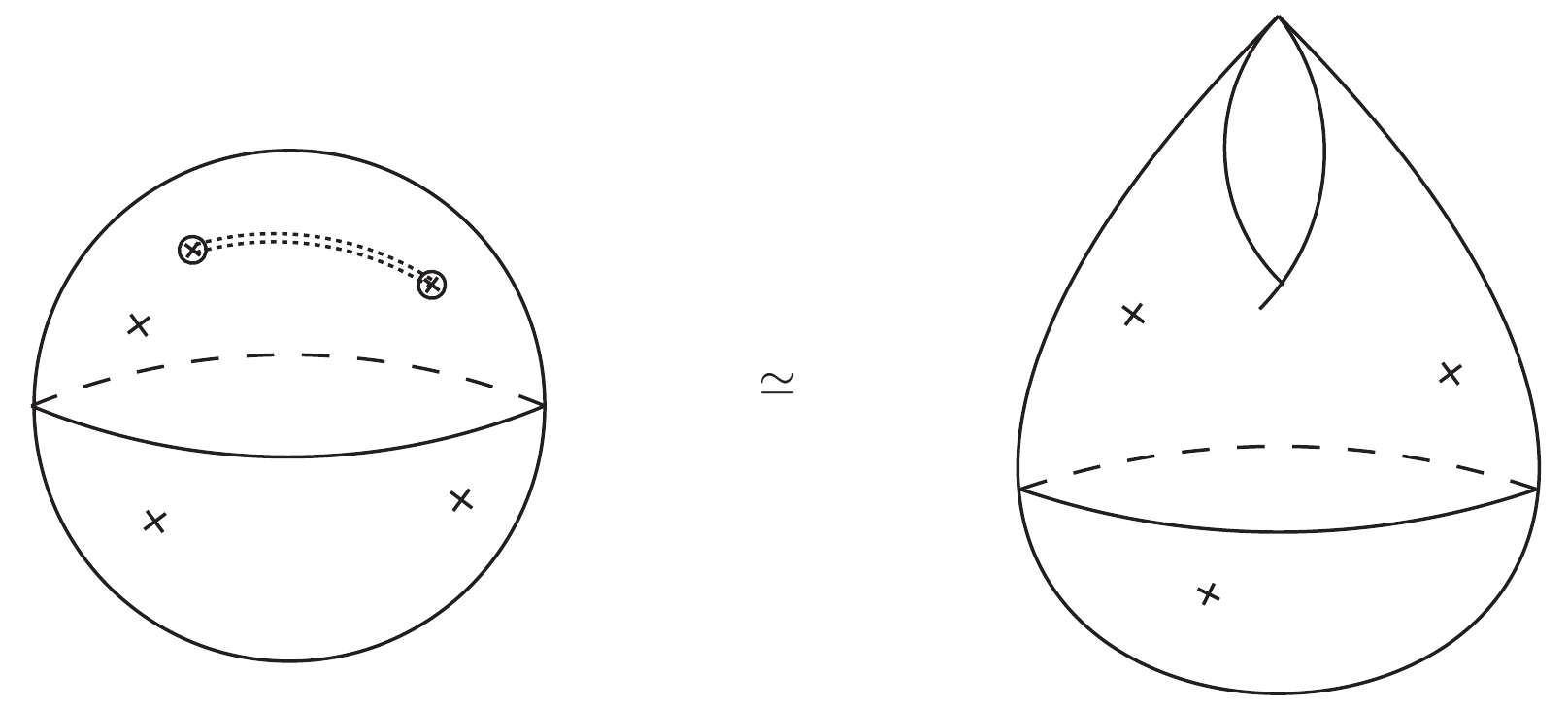}
 \caption{The gluing operator identifies two points in a genus zero correlator, forming a nodal sphere.}
 \label{fig:Gluing2}
 \end{figure}

Similarly, we claim that the 1-loop integrands of \cite{Geyer:2015bja} are in fact correlators of the $g=0$ ambitwistor string CFT computed by inserting the gluing operator with both `legs' on the same sphere, in addition to the usual vertex operators \cref{eqn:LocaVertexOps}. Schematically, we can express this as
 \begin{equation}
 \int _{\overline{\mathcal{M}}_{1,n}} \left\langle   \,  \mathcal{O}_1(z_1) \cdots  \mathcal{O}_n(z_n) \, \right\rangle _\Sigma ~=~   \int _{  \overline{\mathcal{M}}_{0,n+2}} \left\langle   \,    \Delta (z_+, z_- ) ~     \mathcal{O}_1(z_1) \cdots  \mathcal{O}_n(z_n)   \,  \right\rangle _{\Sigma}  
 \end{equation}
 where and $z_\pm$ the locations of the nodes. (See~\cref{fig:Gluing2}.)
 
\medskip

The plan of this paper is as follows. In section~\ref{sec:review} we review a few salient points in the construction of the ambitwistor string, highlighting the form of the BRST operator that will be useful in the rest of the paper. Section~\ref{sec:CubicScalar} gives a detailed description of the gluing operator in bi-adjoint $\phi^3$ theory at tree--level, showing how it can be used to reconstruct tree amplitudes in CHY form. We move to demonstrate how its insertion into the $g=0$ ambitwistor string path integral generates the nodal sphere form of the loop integrand found by~\cite{Geyer:2015bja,Geyer:2015jch}, for cubic scalar, Yang-Mills and gravity in sections \ref{sec:CubicScalar}, \ref{sec:YangMills} and \ref{sec:Gravity} respectively. We conclude with open questions in section~\ref{sec:Conclusions}.

 \section{A brief review of the ambitwistor string} 
  \label{sec:review}
  
  Let us begin by briefly reviewing the salient points in the construction of the ambitwistor string. Further details may be found in~\cite{Mason:2013sva,Adamo:2013tsa,Ohmori:2015sha}. All ambitwistor strings are based on the chiral, bosonic action
 \begin{equation}\label{eqn:actionPreGaugefixing}
 S = \int _\Sigma P \, \partialb X - \frac{e}{2} \, P^2 
 \end{equation}
where $X:\Sigma\to M$ and $P\in\Omega^{1,0}(\Sigma,T^*M)$. For the purposes of scattering theory we take $M=\mathbb{C}^D$ to be (the complexification of) $D$-dimensional flat space, though the ambitwistor string can also be placed on a curved background~\cite{Adamo:2014wea}. The field $e\in\Omega^{0,1}(\Sigma,T_\Sigma)$  transforms like a Beltrami differential on $\Sigma$ and acts as a Lagrange multiplier enforcing the constraint 
 \begin{equation}\label{eqn:Tree-ScattEqns-Sphere}
 P^2(z) = 0 ~, \qquad \forall \, z \in \Sigma ~,
 \end{equation}
so that $P_\mu$ must be null. This constraint generates the gauge transformations
\begin{equation}\label{eqn:BosonicGauge}
X^\mu \mapsto X^\mu + \alpha\, \eta^{\mu\nu} P_\nu \qquad\qquad
e \mapsto e + \partialb \alpha
\end{equation}
where $\alpha$ is a smooth (1,0)-vector on $\Sigma$, whilst $P_\mu$ itself remains invariant. Thus $X$ is defined modulo translation along lightlike directions and, accounting for this redundancy, the moduli space of the ambitwistor string is the space of light rays in $M$, known as 
{\it ambitwistor space}. 

Aside from the action, the only other occurrence of the field $X(z)$ comes from the vertex operators. For plane waves of momentum $p_i$, the $X$ path integral is thus
\begin{equation}
 \int \mathcal{D}X ~ e^{- \int P \partialb X } ~~ \prod_{i=1}^n e^{i \, p_i \cdot X(z_i)}
 \end{equation}
in the presence of $n$ such vertex operators. Integrating out the zero mode of $X$ leads to a momentum--conserving $\delta$-function, whilst integrating out the non-zero modes of $X$ freezes the quantum field $P(z)$ to its classical value
 \begin{equation}\label{eqn:PClassicalTree}
   P(z) =  \sum_{i=1}^n \, p_i \, \omega_{i \ast}(z)   ~,
 \end{equation}
where
\begin{equation}\label{eqn:HolomorphicOneforms-Definition}
 \omega_{i\ast} (z) \equiv \frac{\diff z ~ (z_i - z_\ast)}{(z-z_i)(z-z_\ast)}
 \end{equation}
is the unique meromorphic one-form on the Riemann sphere with simple poles at $z_i$ and $z_\ast$, with residues $\pm1$, respectively. By overall momentum conservation, $P(z)$ in~\eqref{eqn:PClassicalTree} is in fact independent of the auxiliary reference point $z_\ast$. (The introduction of $z_\ast$ is required to make the kinetic operator of the $P,X$ system invertible and thus define a propagator.)

As explained in~\cite{Mason:2013sva}, the vertex operators lie in the BRST cohomology only if $p_i^2=0$ , so the  constraint~\eqref{eqn:Tree-ScattEqns-Sphere} becomes
\begin{equation}
	P^2(z) = 2\sum_{i\neq j} p_i\cdot p_j \,\omega_{i\ast}(z)\,\omega_{j\ast}(z) = 0
\end{equation}
which constrains the location of the punctures in terms of the external momenta, i.e., it says that the worldsheet CFT correlator is only supported at certain special points in $\overline{\mathcal{M}}_{0,n}$. Since any meromorphic quadratic differential on a Riemann sphere must have at least 4 poles (counted with multiplicity), the requirement that $P^2(z)=0$ throughout $\Sigma$ can be enforced by asking
 \begin{equation}\label{eqn:Tree-ScattEqns-Sphere-3}
 \text{Res}_{z=z_i} \, P^2( z )  ~=~ 2 \,  \sum_{j \neq i} p_i \cdot p_j   ~   \omega_{j \ast}(z_i) = 0 
 \end{equation}
at (any) $n-3$ of the $n$ punctures. These are the {\it scattering equations}~\cite{Fairlie:1972zz,Gross:1987kza,Gross:1987ar,Cachazo:2013hca}. They arise in the ambitwistor string from a careful treatment of the moduli of the ghost system associated with the gauge transformation~\eqref{eqn:BosonicGauge}; see~\cite{Adamo:2013tsa,Ohmori:2015sha} for further details. (They can also be seen as arising from the ambitwistor cohomology classes used to represent the external states~\cite{Mason:2013sva}.)

 \subsection{CHY formulae from the ambitwistor string}
 
 In fact, by itself, the bosonic theory above suffers from various anomalies, as may be expected since the worldsheet theory is chiral. One possibility to cure these is to add two sets of fermions, {\it both} of which are left--moving:
$$
 	\psi\in \Omega^0(\Sigma,K^{1/2}_\Sigma\otimes TM) \qquad \text{and} \qquad
	\tilde\psi\in \Omega^0(\Sigma,K^{1/2}_\Sigma\otimes TM)~.
$$
These fields have exactly the same worldsheet quantum numbers, but are subject to independent GSO projections. These fields have action
\begin{equation}\label{eqn:psisystem}
	S[\psi,\tilde\psi] = \int_\Sigma \tfrac{1}{2} \, \psi\cdot\partialb\psi + \tfrac{1}{2} \, \tilde{\psi}\cdot\partialb\tilde{\psi} - \chi\, P\cdot\psi - \tilde\chi \,P\cdot\tilde\psi
 \end{equation}
where the fields $\chi,\tilde\chi\in\Omega^{0,1}(\Sigma,T_\Sigma^{1/2})$ impose constraints $P\cdot\psi=P\cdot\tilde\psi=0$. This ambitwistor string is thus very similar to the RNS superstring, but note that {\it i)} the theory is chiral, with both sets of fermions living in $K^{1/2}_\Sigma$ and {\it ii)} whilst the algebra of constraints $\{P^2, P\cdot\psi, P\cdot\tilde\psi\}$ is similar to that of worldsheet supersymmetry in the RNS string, here $P$ is independent of $X$ (in particular, $P\neq \partial X$) so the transformations they generate have nothing to do with worldsheet diffeomorphisms; rather, the model is a form of worldsheet gauge theory with gauge supergroup $PSL(1|1;\mathbb{C})$. (A pure spinor version of the ambitwistor string has been constructed in~\cite{Berkovits:2013xba}.)

 We BRST quantize this theory by introducing ghost/anti-ghost pairs, along with Nakanishi-Lautrup fields $ H ,  G , \tilde G$, for each of the gauge symmetries and adding the gauge fixing term 
 \begin{equation}
 S_{\text{FP}}  = \int_\Sigma \big\{ Q \, ,~ \tilde b \, (e - e_0) + \beta \, ( \chi - \chi_0)  + \tilde\beta \, ( \tilde\chi - \tilde\chi_0) \big\}
 \end{equation}
to the action, where $e_0 , \chi_0 , \tilde\chi _ 0 $ are the coordinates on moduli space, i.e. the directions of field space transverse to the gauge orbits. The moduli space is finite dimensional and is the remainder of the original path integral after gauge fixing. The BRST operator $Q$ is defined to act as
 \begin{equation}\label{eqn:BRSTaction}
 \begin{aligned}
 &Q \circ (  X  \, ,~ P)   = (  \tilde c \, P + \gamma \, \psi + \tilde\gamma \, \tilde\psi \, ,~ 0 ) ~,\\
 &Q \circ (  \psi  \, ,~ \tilde\psi)   =  ( -  \gamma \, P \, ,~ - \tilde\gamma \, P ) ~,\\
 &Q \circ e  = \partialb \tilde c - 2\, \gamma \, \chi - 2 \, \tilde\gamma \, \tilde\chi ~, \\
 &Q \circ (  \chi  \, ,~ \tilde\chi)   =  (  \partialb \gamma  \, ,~   \partialb \tilde\gamma  ) ~,\\
 &Q \circ \tilde c   =  \gamma^2+\tilde\gamma^2  ~, \qquad Q \circ ( \gamma  \, , ~ \tilde\gamma ) = 0 ~,  \\
 &Q \circ \tilde b   =  H ~, \qquad ~~~ Q \circ ( \beta  \, , ~ \tilde\beta ) = ( G       \, , ~ \tilde G    )   ~, \\
 &Q \circ (H, G , \tilde G) = 0 ~.
 \end{aligned}
 \end{equation}
To obtain the correct measure on moduli space we let $Q$ act as the exterior derivative on moduli space \cite{Witten:2012bh}
 \begin{equation}
 Q \circ e_0    =  \diff e_0 ~, \qquad Q \circ ( \chi_0  \, , ~ \tilde\chi_0 ) = ( \diff \chi _0 \, , ~ \diff \tilde\chi_0)  ~.
 \end{equation}
 Note that this assigns ghost number one and odd Grassmann statistics to the exterior derivative. For the purpose of doing computations using the unfixed action it is convenient to split it into a free-, interaction-, and moduli-part as
 \begin{equation}\label{eqn:GaugeFixedActionComplete}
 \begin{aligned}
  S_{1} &=  \int _\Sigma P \, \partialb X   + \tfrac{1}{2} \,  \psi \, \partialb \psi   + \tfrac{1}{2} \,  \tilde \psi \, \partialb \tilde \psi  - \tilde  b \, \partialb \tilde c  + \beta \, \partialb \gamma + \tilde  \beta \, \partialb \tilde \gamma + H \,e  +  G \, \chi + \tilde G \, \tilde  \chi     \\
   S_{2} &=  \int _\Sigma  -  \frac{e}{2} \, P^2 -  \chi \, \psi \cdot P -  \tilde \chi \, \tilde  \psi \cdot P +2 \, \tilde b \, \gamma \chi +2  \, \tilde b \, \tilde \gamma \tilde \chi  \\
 S_{3} &=  \int _\Sigma   - H \, e_0 + \tilde b \, \diff e_0 - G \, \chi_0- \beta \, \diff \chi_0 - \tilde G \, \tilde{\chi}_0 - \tilde \beta \, \diff \tilde{\chi}_0 
 \end{aligned}
 \end{equation}
 and treat the interactions perturbatively when necessary. Following standard BRST quantization, the BRST charge $Q$ is given by
 \begin{equation}\label{eqn:BRSTproto}
 Q  = \oint c \, T   + \tilde c \, H + \gamma \, G + \tilde \gamma \, \tilde  G  - ( \gamma ^2 + \tilde \gamma ^2)  \, \tilde b~,
 \end{equation}
 and the action on the fields \cref{eqn:BRSTaction} follows from the propagators given by $S_1$ as well as the interactions provided by $S_2$. This BRST operator obeys $Q^2=0$ quantum mechanically iff the target space $M$ has $D=10$.

 Altogether, after integrating out the gauge fields, the \emph{effective} BRST operator for this ambitwistor string is
 \begin{equation}\label{eqn:effectiveBRSTstandard}
 Q_{\rm eff} = \oint c\,T + \frac{\tilde c}{2} \, P^2 +  \gamma\, \psi \cdot P + \tilde\gamma \, P\cdot \tilde\psi    -  (\gamma^2 + \tilde{\gamma}^2) \, \tilde b
 \end{equation}
where $T = P\cdot\partial X + \cdots$ is the worldsheet stress tensor, $c$ is the usual ghost for worldsheet diffeomorphisms, and $\tilde c, \gamma,\tilde\gamma$ are ghosts associated to the symmetries generated by the constraints above. Later in the paper we will find that the presence of the gluing operator alters the form of this effective BRST operator; we would like to emphasize however that the original form of the operator~\eqref{eqn:BRSTproto} remains valid throughout.

In~\cite{Mason:2013sva,Adamo:2013tsa} it was shown that the BRST cohomology consists of only massless states. Fixed vertex operators in the BRST cohomology and surviving the GSO projections take the form
 \begin{subequations}\label{eqn:LocaVertexOps}
 \begin{equation}
 \mathcal{O}(z) = c(z) \tilde{c}(z) \, V(z)  \tilde{V}(z) \, e^{i \, p \cdot X(z)}~,
 \end{equation}
 where 
 \begin{equation}
 V_\text{NS} (z) = \delta(\gamma(z) )  ~ \varepsilon \cdot \psi(z)  
 \end{equation}
 in the Neveu-Schwarz sector and
 \begin{equation}
 V_\text{R}(z) = e^{-\phi(z)/2} \, \zeta^\alpha \, \Theta_\alpha (z)  
 \qquad \text{or} \qquad    
 V_\text{R} (z) = e^{-\phi(z)/2} \, \bar{\zeta}_{\dot \alpha} \, \Theta^{\dot \alpha} (z)  
 \end{equation}
  \end{subequations}
 in the Ramond sector. Here, $\varepsilon_\mu$ is a polarization vector while $\zeta , \bar \zeta$ are left/right handed polarization spinors\footnote{We adopt the conventions of Wess \& Bagger for $10d$ chiral Dirac matrices
$$
 \gamma^\mu_{ \alpha \dot \beta} ~,~~  (\bar \gamma^\mu)^{ \dot \alpha  \beta} ~,~~ C\indices{^{\dot \alpha}_\beta} = -(C^T)\indices{_\beta^{\dot \alpha}}  ~,~~ (C^{-1})\indices{^\alpha_{\dot \beta}}  = - ({C^{-1}}^T)\indices{_{\dot \beta}^\alpha} 
$$
 and the chiral spin fields $\Theta_\alpha (z), \Theta^{\dot \alpha}(z)$. Further, we use the common notation $p_{\alpha \dot \beta } = p_\mu \, \gamma^\mu _{\alpha \dot \beta }$ and $\bar{p}^{\dot \alpha  \beta } = p_\mu \, (\bar{\gamma}^\mu)^{\dot \alpha  \beta }$ for mapping a vector to a bispinor. The Dirac matrices have a number of important properties, for instance
$$
(p \, C)^T = {}-  p \,C  \qquad \text{and} \qquad  (  C^{-1} \, \bar p )^T = {}- C^{-1} \, \bar p  ~.
$$
 These properties will be useful later.}, and $\phi(z)$ is part of the bosonization of the $\beta\gamma$ ghost system (see e.g. \cite{DHoker:1988pdl} for details). Note that the spin fields $\Theta(z)$ carry holomorphic conformal weight $\sfrac{5}{8}$ while $e^{ q \phi }$ carries holomorphic conformal weight $- q \,(q/2 +1)$. The OPEs among spin fields and fermions are the same as in usual string theory, namely
\begin{equation}\label{eqn:SpinfieldOPEs}
\begin{aligned}
\Theta_\alpha (z) ~ \Theta_\beta(w) & ~\sim~  S(z,w)^{1/4} \sqrt{ \omega_{w \ast }(z)} ~ (\gamma^\mu C)_{\alpha\beta}  \, \eta_{\mu\nu}~ \psi^\nu (w) \\
\Theta^{\dot \alpha} (z) ~ \Theta_\beta(w) & ~\sim~ S(z,w)^{5/4} ~ C\indices{^{\dot \alpha}_\beta}  \\
\psi^\mu (z) ~ \Theta_\alpha (w) & ~\sim ~ \sqrt{ \omega_{w \ast }(z)} ~ \gamma^\mu_{\alpha \dot \beta } \, \Theta^{\dot \beta} (w)  \\
\psi^\mu (z)  ~ \Theta^{\dot \alpha } (w) & ~\sim ~ \sqrt{ \omega_{w \ast }(z)} ~ (\bar {\gamma}^\mu)^{\dot \alpha  \beta } \, \Theta_{ \beta} (w) \\
\psi^\mu (z)  ~ \psi^\nu (w) & ~\sim ~ S(z,w) \, \eta^{\mu\nu} \\
e^{q \, \phi(z)} ~ e^{q^\prime \, \phi(w)} &~\sim~  (z-w)^{-q q^\prime} ~ e^{(q + q^\prime) \, \phi(w)}   ~.
\end{aligned}
\end{equation}
In principle, these may be used to evaluate any worldsheet correlator in the Ramond sector.  There are analogous expressions for the spin fields in $\tilde V$, which again we emphasize also carry {\it holomorphic} conformal weight.  After imposing a GSO projection, the spectrum is that of $D=10$ Type II  A/B supergravity~\cite{Adamo:2013tsa}, according to whether the untilded and tilded Ramond sectors are chosen to have opposite/same space-time chiralities.

\bigskip

The CHY formula~\cite{Cachazo:2013hca} for $n$-particle tree-level scattering amplitudes in gravity follows from the genus zero correlator of $n$ vertex operators in the NS sector, together with the appropriate integrals over the (bosonic and fermionic) worldsheet moduli space. In particular, each of the two CHY Pfaffians arises from the correlator of the fermions in the $n$ vertex operators, together with the fermions in the $n-2$ picture changing operators that can be viewed as coming from integrating over the fermionic moduli space. Thus, for the $\psi$s, we have
 \begin{equation}\label{eqn:PsiCorrelator-Explanation}
 \left\langle \prod_{i=1}^n \varepsilon \cdot \psi(z_i)   ~ \prod_{r=1}^{n-2}  \psi(x_r) \! \cdot \! P(x_r)    \right\rangle = \pf \left(    \begin{array}{cc}
A & -C^T \\ C & B 
\end{array}     \right)
 \end{equation}
 with the sub-matrices given by\footnote{$S(z_i,z_j)$ is the genus zero Szeg\'o kernel
$$
	S(z_i,z_j) \equiv \frac{\sqrt{\diff z_i\,\diff z_j}}{z_i-z_j}~.
$$} 
 \begin{equation}
 \begin{aligned}
 A_{rs} &= P(x_r) \cdot P(x_s) \, S(x_r , x_s)  \qquad &\text{for } r  & \neq s ~, \qquad &A_{rr} &= 0  ~, \\
 B_{ij} &= \varepsilon_i \cdot \varepsilon_j \, S(z_i , z_j)  \qquad &\text{for } i & \neq j ~, \qquad  & B_{ii} &= 0  ~, \\
 C_{ir} &= \varepsilon_i \cdot P(x_r) \, S(z_i , x_r)  ~. && &&
 \end{aligned}
 \end{equation} 
We recall that the $X$ path integral has frozen the field $P(z)$ to its classical value \eqref{eqn:PClassicalTree}. This fermion correlator is accompanied by the correlator of the ghost insertions
 \begin{equation}\label{eqn:BetaGammaCorrelator-Explanation}
 \left\langle \prod_{i=1}^n \delta(\gamma(z_i))   ~ \prod_{r=1}^{n-2}  \delta(\beta(x_r)) \,   \right\rangle  = \frac{\prod_{i<j=1}^n  S(z_i,z_j)  \prod_{r<s=1}^{n-2}  S(x_r , x_s) }{\prod_{i=1}^n \prod_{r=1}^{n-2} \, S(z_i,x_r)}
 \end{equation}
 and, on the support of the scattering equations, Liouville's theorem shows that the product of \eqref{eqn:PsiCorrelator-Explanation} \& \eqref{eqn:BetaGammaCorrelator-Explanation} is independent of the locations $x_r$, as expected for PCOs. In particular, we can take the auxiliary points $x_r$ to coincide with $n-2$ of the external punctures, upon which we recover the original Pfaffian of CHY~\cite{Cachazo:2013hca}. One can take this limit already at the level of the CFT correlator, changing $n-2$ of the picture number $-1$ vertex operators to 
 \begin{equation}
	\lim_{x \to z} \left( \delta(\beta)\, P \cdot \psi \right) \!(x)  ~  \left( \delta(\gamma) \, \varepsilon \cdot  \psi\,e^{ip\cdot X} \right) \! (z)
	 =  \left( (\varepsilon \cdot P + \varepsilon \cdot \psi \,   p \cdot \psi ) \, e^{ip\cdot X} \right) \! (z)
 \end{equation}
of picture number 0. A similar correlation function for the $\tilde\psi$ system produces the second Pfaffian in the CHY formula for gravity amplitudes.

\medskip

At present, there is no known, consistent ambitwistor string describing Einstein--Yang--Mills theory (nor pure Yang--Mills theory). However, one can generate CHY formulae for tree amplitudes in these theories from anomalous ambitwistor strings, provided one is willing to discard certain terms `by hand'. For example, if we replace the $\tilde\psi$ fermion system by a general worldsheet current algebra $j^a(z)$ obeying
  \begin{equation}
  j^a (z) \, j^b(0) \sim \frac{k }{z^2}  \, \delta^{ab}+ \frac{1}{z}  \, f^{ab}_c \, j^c(0) + \order {z} 0 ~,
  \end{equation}
 then the NS vertex operator
\begin{subequations}
\begin{equation}
 	\mathcal{O}_{\rm NS}(z) = c(z)\tilde c(z) V_{\rm NS}(z)\, t_a j^a(z)\,e^{ip\cdot X(z)}
\end{equation}
describes a gluon of polarization  $\epsilon_\mu$ and colour $t_a$, while R sector vertex operator
\begin{equation}\label{eqn:GluinoVertex}	
	\mathcal{O}_{\rm R}(z) = c(z)\tilde c(z) V_{\rm R}(z)\, t_a j^a(z)\,e^{ip\cdot X(z)}
\end{equation}
\end{subequations}
describes a gluino of spin $\zeta$ and colour $t_a$.  As shown in~\cite{Mason:2013sva}, inserting $n$ gluon vertex operators and keeping only the leading trace contribution yields the CHY formula for tree-level scattering in Yang-Mills theory~\cite{Cachazo:2013hca}, with one Pfaffian arising from the $\psi$s as before, and the leading--trace current correlator giving a worldsheet Parke--Taylor factor
\begin{equation}\label{eqn:ParkeTaylorDefn}
  \text{PT}(\alpha) = \tr\left( t_{\alpha(1)}  \cdots t_{\alpha(n)}  \right) ~ \prod_{i=1}^n S(z_{\alpha(i)} , z_{\alpha(i+1)}) ~,
  \end{equation}
summed over all inequivalent colour orderings $\alpha$, and indices taken modulo $n$. Similarly, replacing both $\psi$ and $\tilde\psi$ by two, independent current algebras $j_a(z)$ and $\tilde\jmath_{\tilde a}(z)$ and inserting vertex operators 
\begin{equation}\label{eqn:BiAdjointScalarVertex}
	\mathcal{O}(z) = c(z)\tilde c(z)\,t_aj^a(z)~{\tilde t}_{\tilde a}\tilde\jmath^{\tilde a}(z)~e^{ip\cdot X(z)}\,,
\end{equation}
the (double) leading trace terms yield the CHY formula for the cubic, bi-adjoint scalar theory, with two Parke--Taylor factors summed over independent colour orderings $\alpha$ and $\beta$:
\begin{equation}
	\mathcal{M}_{\text{bi-adjoint}} = \delta^D\!\left(\sum_{i=1}^n p_i\right)\,\sum_{\alpha,\beta\in S_n/\mathbb{Z}_n} \mathfrak{m}(\alpha,\beta)~,
\end{equation}
where the colour--ordered partial amplitudes 
 \begin{equation}\label{eqn:PhicubedTreeAmplitude}
  \mathfrak{m}(\alpha , \beta) = \int \left(\frac{1}{ \omega_{123}} \right)^2  \,  \prod_{i = 4}^n \deltab \left( \res _{z_i} P^2 \right)  ~ \text{PT}(\alpha) ~ \text{PT}(\beta)~,
  \end{equation}
 and $\omega_{123} = S(z_1,z_2) \, S(z_2,z_3) \,  S(z_3,z_1)$ is the M{\"o}bius volume factor.

\medskip

We stress that, unlike the Type II gravity ambitwistor string, the (known) models containing such current algebras are not consistent. Nonetheless, we will find it convenient to consider such anomalous models below, so as to illustrate the gluing operator in a simpler context.


\section{The gluing operator for the bi-adjoint scalar}\label{sec:CubicScalar}
We now begin our presentation of the detailed form of the gluing operator $\Delta(z_\ast,z'_\ast)$. In this section we will consider the gluing operator for the bi-adjoint scalar theory, demonstrating its use in computing tree and 1-loop amplitudes, each in a given colour ordering. As explained in the introduction, for tree amplitudes, $z_\ast$ and $z'_\ast$ should be chosen to be points on separate Riemann spheres, joined by this operator, while for the 1-loop amplitude they will both be inserted on the same Riemann sphere. 
   
\bigskip

The gluing operator for the bi-adjoint scalar theory takes the form 
 \begin{equation}\label{eqn:GluingBiAdjoint1}
    \Delta_{\phi^3}(z_\ast,z'_\ast) 
    = \int \frac{\diff^d \ell}{\ell^2} ~ W_{ij}(z_\ast) \, W_{kl}(z_\ast^\prime) ~  \mathcal{O}^{a\tilde a}(z_\ast) ~  \Delta_{ab\,\tilde{a}\tilde{b}} \,      \mathcal{O}^{b\tilde b}(z_\ast^\prime)~,
 \end{equation}
 where $i,j,k,l \in \{1 , \cdots n \} $ are four external punctures with a special role, described below. The operators $\mathcal{O}^{a\tilde a}(z_\ast)$ and $\mathcal{O}^{b\tilde b}(z'_\ast)$ are local insertions on the left/right sphere, respectively, defined by
 \begin{equation}\label{eqn:GluingBiAdjoint2}
 \begin{aligned}
	 \mathcal{O}^{a\tilde a}(z_\ast) &= c \, \tilde{c } ~ j^a \, \tilde{\jmath}^{\tilde a} \, e^{i \, \ell \cdot X}(z_\ast) \\
	  \mathcal{O}^{b\tilde b}(z'_\ast) &= c \, \tilde{c } ~ j^b \, \tilde{\jmath}^{\tilde b} \, e^{-i \, \ell \cdot X}(z'_\ast)~.
\end{aligned}
 \end{equation}
Note the signs of the momenta in the exponentials, corresponding to off-shell momentum $\ell$ flowing into the left Riemann sphere and out of the right. The only differences between $\mathcal{O}^{a\tilde a}$ and the local on-shell vertex operators \eqref{eqn:BiAdjointScalarVertex} are that the momentum $\ell$ in~\eqref{eqn:GluingBiAdjoint2} is not required to be null, and that the operators in~\eqref{eqn:GluingBiAdjoint2} have arbitrary colour. The operators are joined by the tensor structure of the target space Feynman propagator for a scalar
\begin{equation}\label{eqn:GluingBiAdjoint3}
	\Delta_{ab\,\tilde a\tilde b} = \delta_{ab}\,\delta_{\tilde a \tilde b}
\end{equation}
which connects the colour flow from one Riemann sphere to the other\footnote{In \eqref{eqn:GluingBiAdjoint3} we have given the colour structure for a $U(N) $ gauge group. However, since the colour singlet decouples by the Kleiss-Kuijf relations, this also holds for $SU(N)$.}. Thus, this part of the gluing operator represents a set of `off-shell states' flowing through the node.

Since these ingredients describe off-shell states they cannot, by themselves, be invariant under the original BRST operator on each Riemann sphere separately. The failure of $\mathcal{O}^{a\tilde a}(z_\ast)$ to be BRST closed is compensated on each side separately by the remaining ingredient
\begin{equation}\label{eqn:W-FactorTree-cubicscalar}
  W_{ij}(z_\ast)  = \exp \left(  \frac{\ell^2}{2} \int_\Sigma e(x) \, \omega_{i\ast}(x)\, \omega_{j\ast}(x)   \right)
 \end{equation} 
which is a non-local operator. $W_{ij}(z_\ast)$ depends on $z_\ast$ (as well as $z_i,z_j$) through the meromorphic 1-forms $\omega_{i\ast}$ and $\omega_{j\ast}$. Crucially, $W_{ij}(z_\ast)$ carries dependence on the gauge field $e$, and hence will modify effective BRST operator to 
\begin{equation}
 Q_{\rm eff} = \oint c\,T + \frac{\tilde c}{2} \, \left(P^2 - \ell^2\,\omega_{i\ast}\,\omega_{j\ast} \right)
 \end{equation}
on the left Riemann sphere, with a similar effective BRST operator on $\Sigma_R$. We emphasize that the fundamental BRST operator~\eqref{eqn:BRSTproto} remains unchanged; this modified \emph{effective} BRST operator emerges naturally after integrating out the gauge fields and applying standard BRST quantization in the presence of the gluing operator.


\subsection{Tree amplitudes in $\phi^3$ theory}

Using this gluing operator, we claim that at tree-level the colour--ordered partial amplitudes $\mathfrak{m}(\alpha,\beta)$ in the cubic bi-adjoint scalar theory can be written as
 \begin{equation}\label{eqn:SumOverChannels}
 	\mathfrak{m}(\alpha , \beta ) =  \sum_{\substack{   \alpha_L \,  \cup  \, \alpha_R  = \alpha \\  \beta_L \,  \cup  \, \beta_R  = \beta  \\ \text{compatible} }} \bar{\mathfrak{m}}^{a\tilde a} (\alpha_L , \beta_L )  ~ \Delta_{ab\, \tilde a\tilde b} ~ \bar{\mathfrak{m}}^{b\tilde b} (\alpha_R , \beta_R ) 
 \end{equation}
where $\bar{\mathfrak{m}}^{a\tilde a} (\alpha_L ,\beta_L) $ denotes the correlator
\begin{equation}\label{eqn:mBarBiAdjoint}
	\bar{\mathfrak{m}}^{a\tilde a} (\alpha_L ,\beta_L) 
	= \int_{\mathcal{M}_{0,n_{L}+1}}  \left.   \left\langle    W_{ij} (z_\ast) ~ \mathcal{O}^{a\tilde a}(z_\ast) ~ 
	\prod_{i\in L} \mathcal{O}_i (z_i) \right\rangle   \right|_{\alpha_L , \beta_L} ~,
\end{equation}
involving insertions of one leg of the gluing operator, together with the vertex operators for the particles that are on the `left' Riemann sphere  (similarly for $\bar{\mathfrak{m}}^{b\tilde b} (\alpha_R ,\beta_R)$). The original $n$-particle colour ordering $\alpha$ has split into two $\alpha_{L/R} \in S_{n_{L/R}+1}$, where the extra element is the node, and likewise for $\beta$. (The subscripts on the correlator instruct us to take just these colour orderings.) The split of particles into left and right has to be \emph{compatible} with the original colour ordering, in the sense that stepping through the entire set, from $\alpha(i)$ to $\alpha(i+1)$, must require crossing from the left to the right sphere, and back, exactly once. The sum over channels means a sum over all ways of splitting the external particles into two ordered sets, compatible with colour ordering in the way just described, and with the property that exactly two out of the four particles $i,j,k,l$ are on each side. We have to make a choice of $i,j,k,l$, but the sum is independent of this choice - as we explain below.

\medskip

The correlator~\eqref{eqn:mBarBiAdjoint} represents a partial amplitude with one leg off-shell. It can be easily evaluated. Integrating out $X$ from the Riemann sphere on the left, we find as usual a momentum--conserving $\delta$-function that sets 
\begin{equation}\label{eqn:MomFixesEll}
	\ell+\sum_{i\in L}p_i =0~,
\end{equation}
and that 
 \begin{equation}
 	P_{L}(z) = \sum_{i \in \mathrm{L}} p_i ~ \omega_{i\ast}(z) ~.
 \end{equation}
 Here, we have used~\eqref{eqn:MomFixesEll} to eliminate the explicit dependence of $\ell$ from $P_L(z)$; note that it is now important that each term in the sum has a simple pole at the node $z_\ast$.
 
The presence of the gauge field $e$ in the gluing operator alters the scattering equations. Integrating it out, we now find that $\bar{\mathfrak{m}}$ on the left is supported on solutions of the `off-shell scattering equations'
\begin{equation}\label{eqn:off-shell-scatteqns}
\begin{aligned}
P_L^2(z) &= \ell^2 ~  \omega_{i\ast}(z) \, \omega_{j\ast}(z)  \\
\end{aligned}
\end{equation}
Similarly, we obtain from the right Riemann sphere a momentum conserving $\delta$-function fixing $-\ell + \sum_{i\in  \mathrm {R}}p_i =0$, which together with~\eqref{eqn:MomFixesEll} implies overall momentum conservation, as well as the constraints
\begin{equation}
P_R^2(z') =  \ell^2   ~\omega_{k\ast}(z') \, \omega_{l\ast}(z') 
\qquad\text{with}\qquad
P_R(z') = \sum_{i\in \mathrm{R}} p_i~\omega_{i\ast}(z')
\end{equation}
on the right hand sphere. Note that these conditions, together with the definition of $P_L(z)$, are exactly what we would obtain in the limit that the original Riemann sphere degenerates to form a node. 

After performing the CFT path integral to compute the correlator, we find explicitly
\begin{equation}
\bar{\mathfrak{m}}^{a\tilde a} (\alpha_L ,\beta _L ) = \int \left( \frac{1}{\omega_{ij\ast}} \right)^2 \prod_{i \in L \backslash \{i,j\} } \deltab (\res_{z_i}P_L^2) ~~ \text{PT}(\alpha_L)^a ~   \text{PT}(\beta_L)^{\tilde a}
\end{equation}
for the left building block, and similarly for the right. The Parke-Taylor factors with an index are defined as
 \begin{equation}\label{eqn:ParkeTaylorIndex}
 \text{PT}(\alpha)^a = \tr\left( t_{\alpha(1)}  \cdots t_{\alpha(n)} \, t^a  \right) ~ \prod_{i=1}^{n-1} S(z_{\alpha(i)} , z_{\alpha(i+1)}) ~ S(z_{\alpha(n)} , z_\ast) ~ S(z_\ast , z_{\alpha(1)}) ~,
  \end{equation}
 which is the straightforward generalization of \cref{eqn:ParkeTaylorDefn}, with the node inserted as a new member of the colour trace. As required by unitarity, when the momentum flowing through the node goes on-shell, the $\bar{\mathfrak{m}}$ become proper on-shell amplitudes, as the gluing operator simply turns into a pair of on-shell vertex-operators traced over the on-shell Hilbert space.

 \medskip
 
 Let us now explain why, with these ingredients, the sum~\eqref{eqn:SumOverChannels} indeed agrees with the original $n$-partial colour ordered amplitude $\mathfrak{m}(\alpha,\beta)$ in~\eqref{eqn:PhicubedTreeAmplitude}. We begin by integrating~\eqref{eqn:PhicubedTreeAmplitude} by parts in the moduli space $\overline{\mathcal{M}}_{0,n}$, and take for concreteness $i,j,k,l = 1,2,3,4$. Explicitly, we write
    \begin{equation}\label{eqn:PhicubedTreeAmplitude-IPB-1}
    \begin{aligned}
  \mathfrak{m}(\alpha , \beta) &=  \int \left( \frac{1}{\omega_{123} } \right)^2  \, \partialb \frac{1}{\res _{z_4}P^2} \,  \prod_{i = 5}^n \deltab \left( \res _{z_i} P^2 \right)  ~ \text{PT}(\alpha) ~ \text{PT}(\beta)\\
  &=-\int \frac{1}{\res _{z_4}P^2} \, \partialb\left[\left( \frac{1}{\omega_{123} } \right)^2  \,    \prod_{i = 5}^n \deltab \left( \res _{z_i} P^2 \right)  ~\text{PT}(\alpha) ~ \text{PT}(\beta)\right]
  \end{aligned}
  \end{equation}
  so that the $\partialb$ derivative originally acting on $\res_{z_4}P^2$ now picks up contributions from any poles in the rest of the integrand. (Equivalently, treating the original integrand as a top meromorphic form on $\overline{\mathcal{M}}_{0,n}$,  we deform the contour, originally surrounding all scattering equations, away from the pole at $\res_{z_4}P^2 =0$ and thus pick up residues from all other poles outside the original contour.) As shown in~\cite{Dolan:2013isa}, the only poles of the expression in square brackets on~\eqref{eqn:PhicubedTreeAmplitude-IPB-1} lie on the boundary of the moduli space $\overline{\mathcal{M}}_{0,n}$ where the Riemann sphere degenerates to a nodal curve with two components. (Recall that we continue to impose the scattering equations for particles $5,\ldots,n$.) Furthermore, we obtain at most a simple pole on boundary divisors that are compatible with the colour orderings $(\alpha,\beta)$ in the sense given above, and in which exactly two of the distinguished points $z_1,z_2,z_3,z_4$ lie on each curve component.

 \medskip
 
It is worth emphasizing that the ambitwistor correlation functions are independent of the choice of $n-3$ points at which we choose to impose scattering equations only as long as \emph{all} the scattering equations are enforced. In order to employ the global residue theorem, we have to {\it choose} a meromorphic form that extends the original CHY integrand off the support of the scattering equations; the form given in the second line of~\eqref{eqn:PhicubedTreeAmplitude-IPB-1} certainly achieves this, but it is far from the unique choice. Each such expression has the same residue at the solutions to the scattering equations, but their value/residues may differ significantly everywhere else. Furthermore, as soon as we relax one of the scattering equations, \eqref{eqn:PhicubedTreeAmplitude-IPB-1} ceases to be independent of the choice of which points have, or do not have, scattering equations imposed. Thus, as soon as we integrate by parts away from the pole at $\res_{z_4}P^2$ in \cref{eqn:PhicubedTreeAmplitude-IPB-1}, we loose invariance under the permutations of the external particles; we have singled out the particles at $z_{1,2,3,4}$ as playing a special role.

More generally, we could talk of independence of the choice of basis of $$T^{1,0}\mathcal{M}_{0,n}\cong H^{0,1}(\Sigma,T_\Sigma(-z_1-\cdots-z_n))$$ that we use to describe the $\tilde b$-ghost moduli responsible for imposing the scattering equations. Fundamentally, this is because the independence of the points $w_r$ is a reflection of the gauge invariance under local transformation generated the charge $P^2$; by going away from the locus of the solutions to the scattering equations, we are breaking this gauge invariance. 

\medskip

 Coming back to our example, after relaxing the scattering equation at $z_4$, the field $P(z)$ no longer obeys $P^2(z)=0$ for all $z\in\Sigma$, since $n-4$ scattering equations are not enough to enforce $P^2 =0$ globally.  The remaining scattering equations
\begin{equation}\label{eqn:scattEqnsLessOne}
 \res_{z_i} P^2 = 0 ~, \qquad \text{for } i = 5 , \cdots , n ~,
 \end{equation} 
are however enough to imply that
 \begin{equation}\label{eqn:scattEqnsLessOne-2}
 P^2(z) = \frac{(w-z_1)(w-z_2)(w-z_3)(w-z_4)}{   (z-z_1)(z-z_2)(z-z_3)(z-z_4)}  ~\frac{(\diff z)^2}{(\diff w)^2 } \,  P^2(w)
 \end{equation}
 where $w$ is a fixed, arbitrary point anywhere on the sphere, and the RHS is independent\footnote{In other words, on the support of the scattering equations $\res_{z_i} P^2 =0$ for $i = 5 , \cdots , n$, the combination
  \begin{equation}
 P^2(z) \, \frac{1}{\diff z^2} ~    (z-z_1)(z-z_2)(z-z_3)(z-z_4) 
 \end{equation}
 is independent of $z$, (though still depends on the $z_i$,) which again follows from Liouville's theorem.} of $w$ by Liouville's theorem: it is a scalar function of $w$ with no poles. Recall that the equations \eqref{eqn:scattEqnsLessOne} are to be understood as a constraint on the moduli of the surface. They can be thought of as fixing all but one of the locations $z_i$, up to M\"obius invariance.

Let us consider the behaviour of $P^2(z)$ as we approach a degeneration in which points $1,2$ lie on the `left' component curve while $3,4$ lie on the `right'.  In this limit, \cref{eqn:scattEqnsLessOne-2} becomes
 \begin{equation}\label{eqn:scattEqnsLessOne-3}
 P_L^2(z) = \frac{(w-z_1)(w-z_2)(w-z_\ast)(w-z_\ast)}{   (z-z_1)(z-z_2)(z-z_\ast)(z-z_\ast)}  ~ \frac{(\diff z)^2}{(\diff w)^2 } \,  P_L^2(w) ~, \qquad \text{for } w,z \in \Sigma_L ~,
 \end{equation}
 and similarly for $w,z \in \Sigma_R$. Recall that $z_\ast$ denotes the location of the emergent node. Note that in going from \eqref{eqn:scattEqnsLessOne-2} to \eqref{eqn:scattEqnsLessOne-3}, the RHS remains independent of the arbitrary point $w$ throughout. We can simplify \eqref{eqn:scattEqnsLessOne-3} by taking the limit $w \to z_\ast$, where it becomes
 \begin{equation}\label{eqn:scattEqnsLessOne-4}
 P_L^2(z) =  \ell^2 ~ \omega_{1\ast}(z) \, \omega_{2\ast}(z) ~,
 \end{equation}
 with $\ell = - \sum_{i \in L} p_i = \sum_{i \in R} p_i$ the momentum flowing through the node. Similarly, on the `right' curve component, with local coordinate $z'$, we have
 \begin{equation}\label{eqn:scattEqnsLessOne-5}
 P_R^2(z') = \ell^2 ~  \omega_{3\ast}(z') \, \omega_{4\ast}(z') ~.
 \end{equation}
 These are exactly the off shell scattering equations~\eqref{eqn:off-shell-scatteqns} that arose above from integrating out the gauge field $e$ in the presence of the gluing operator. Furthermore, from~\eqref{eqn:scattEqnsLessOne-5} it follows that on this boundary divisor
 \begin{equation}
 	\frac{1}{\text{Res}_{z_4}P^2} = \frac{1}{\ell^2\,\omega_{3\ast}(4)}~.
\end{equation} 
 We recognize the factor of $1/\ell^2$ as present in the gluing operator, while the factor of $\omega_{3\ast}(4)$ combines with other ingredients to form the volume of the M{\"o}bius group on the right Riemann sphere.
Finally, the limit of each worldsheet Parke--Taylor factors becomes the coloured Parke--Taylor factor~\eqref{eqn:ParkeTaylorIndex} describing the colour flow through the node. We thus see that the sum over all compatible boundary divisors 
 $$
 	 \sum_{\substack{   \alpha_L \,  \cup  \, \alpha_R  = \alpha \\  \beta_L \,  \cup  \, \beta_R  = \beta  \\ \text{compatible} }} \bar{\mathfrak{m}}^{a\tilde a} (\alpha_L , \beta_L )  ~ \Delta_{ab\, \tilde a\tilde b} ~ \bar{\mathfrak{m}}^{b\tilde b} (\alpha_R , \beta_R ) 
$$
given in~\eqref{eqn:SumOverChannels} indeed corresponds to an evaluation of the original colour--ordered amplitude $\mathfrak{m}(\alpha,\beta)$. We emphasize that this sum gives the {\it complete} (colour--ordered) amplitude, not its cuts. Of course, by varying the external momenta and taking the residue as $\ell^2\to0$ we indeed would obtain the cut amplitude in a given channel. We have checked numerically up to six points that this construction indeed coincides with the original amplitude.


\subsection{One loop amplitudes in $\phi^3$ theory}

To obtain 1-loop amplitudes in this bi-adjoint theory we use essentially the same gluing operator as at tree level, but now with both `legs' inserted on the same copy of a Riemann sphere. Specifically, along with the $n$  vertex operators describing external states, we insert
\begin{equation}
	\Delta_{\phi^3}(z_+,z_-) =  \int \frac{\diff^d \ell}{\ell^2} ~  \mathcal{O}^{a\tilde a}(z_+) ~  \Delta_{ab\,\tilde{a}\tilde{b}} \, \mathcal{O}^{b\tilde b}(z_-)~W(z_+,z_-)  
\end{equation}
where $\mathcal{O}^{a\tilde a}$ and $\Delta_{ab\,\tilde{a}\tilde{b}}$ were defined in~\eqref{eqn:GluingBiAdjoint2} and~\eqref{eqn:GluingBiAdjoint3} above, while now
\begin{equation}
	W(z_+,z_-)  = \exp\left(\frac{\ell^2}{2}\int_\Sigma e(x)\, \omega_{+-}^2(x)\right)~.
\end{equation}
in place of~\eqref{eqn:W-FactorTree-cubicscalar}. Note that in both the tree and 1-loop factors, the $W$s can be understood as describing moduli associated to the normal bundle of the boundary divisor. It can also be seen as a holomorphic Wilson line~\cite{Frenkel:2005qk,Mason:2010yk,Bullimore:2011ni}, transporting a frame for the gauge field $e$ from $z_+$ to $z_-$.

Let us first understand the emergence of the one-loop scattering equations. This follows in much the same way as the modification of the scattering equations at tree level. In the presence of the gluing operator, the $PX$ path integral reads
 \begin{equation}
  \int\mathcal{D}{P}\,\mathcal{D}{X}\,  e^{- \int P \partialb X + \frac{e}{2} P^2} ~~ e^{i \, \ell \cdot ( X(z_+) - X(z_-)) + \frac{\ell^2}{2}\int e \, \omega_{+-}^2 }   ~~ \prod_{i=1}^n e^{i \, p_i \cdot X(z_i)}  ~,
 \end{equation}
where the first term is from the action, the second from the gluing operator and the rest from the external vertex operators. As for tree-level amplitudes,  $X$ and $e$ only appear linearly in the exponential and so act as Lagrange multipliers. As always, integrating out the zero-mode of $X$ produces a momentum conserving $\delta$-function constraining $\sum_i p_i=0$, while integrating out the non-zero modes of $X$ freezes the quantum field $P(z)$ to its classical value
 \begin{equation}\label{eqn:OneLoop-FrozenP}
   P(z) = \ell \, \omega_{+-}(z) + \sum_{i=1}^n p_i \, \omega_{i \ast}(z)  \, ,
 \end{equation}
 where $\omega_{+-}(z) = \frac{\diff z  \, (z_+ - z_-)}{(z-z_+)(z-z_-)}$, in analogy to the one-forms $\omega_{i\ast}$ defined in \eqref{eqn:HolomorphicOneforms-Definition}. Similarly, upon integrating out $e$ we discover that the correlators have $\delta$-function support on the solutions to 
\begin{equation}\label{eqn:OneLoop-ScattEqns-Sphere}
 P^2 = \ell^2 \, \omega_{+-}^2~.
 \end{equation}
  Thus, inserting the gluing operator into the genus zero correlation function modifies the scattering equations to become the same `one-loop scattering equations' one obtains \cite{Geyer:2015bja} by localising the genus one correlation function on the boundary at $q=0$.

Although equation \eqref{eqn:OneLoop-ScattEqns-Sphere} is required to hold at every point on the sphere, the holomorphic nature of $P(z)$ actually makes \eqref{eqn:OneLoop-ScattEqns-Sphere} a finite dimensional constraint. Indeed, it is sufficient to pick a set of $n-1$ arbitrary points $\{w_1,w_2,\ldots,w_{n-1}\}\in\Sigma$ and require that
 \begin{equation}\label{eqn:OneLoop-ScattEqns-Sphere-2}
 P^2( w_r ) =  \ell ^2 \, \omega_{+-}^2(w_r)
\end{equation}
at each of these points. Holomorphy of $P(z)$ then ensures that \eqref{eqn:OneLoop-ScattEqns-Sphere} holds globally.  It is worth pointing out that, even though the equations \eqref{eqn:OneLoop-ScattEqns-Sphere-2} individually depend on the choice of points $w_r$, their solutions do not. As a special case, one can take the $w_r$ to coincide with some of the punctures, which, after multiplication with $(w_r-z_i)$, amounts to requiring
 \begin{equation}\label{eqn:OneLoop-ScattEqns-Sphere-3}
 	\text{SE}_i := \text{Res}_{z_i} \bigg\{ P^2( z ) -  \ell ^2 \, \omega_{+-}^2(z) \bigg\} =0
 \end{equation}
 for any $n-1$ of the $n+2$ punctures. 

\medskip

The remaining part of the correlation function comes from the two independent current algebras. To obtain the 1-loop amplitude in a particular colour ordering, we need to extract the coefficient of a given, single-trace contribution {\it by hand}. In addition, we must {\it by hand} only consider contributions in which $z_+$ and $z_-$ are adjacent in the colour ordering, ensuring that the colour `runs around the loop'. These conditions are analogous to the fact that we had to extract single-trace terms by hand even to obtain the tree amplitude. In both cases, they are symptomatic of the fact that this bosonic ambitwistor string model does not correctly describe pure bi-adjoint scalar theory. A further symptom of this sickness is that, as noted in~\cite{Geyer:2015bja,Geyer:2015jch}, the 1-loop scattering equations~\eqref{eqn:OneLoop-ScattEqns-Sphere} contain certain `singular solutions', where $z_+= z_-$, corresponding to a tadpole. These solutions must again be discarded by hand. Later, we will investigate 1-loop amplitudes in gravity using a gluing operator in the Type II ambitwistor strings. In this case, the worldsheet CFT correctly generates the full answer in a consistent manner -- in particular, in this consistent ambitwistor model, it will not be necessary to discard any terms by hand.

\medskip

Altogether, the bosonic ambitwistor string worldsheet correlator of $n$ vertex operators and the gluing operator leads to 
\begin{multline}
	 \int\left\langle \Delta_{\phi^3}(z_+,z_-) \,\prod_{i=1}^n\,
	c(z_i)\,\tilde c(z_i)~t^{(i)}_aj^a(z_i)~{\tilde t}^{(i)}_{\tilde a}\tilde\jmath^{\tilde a}(z_i)~e^{ip\cdot X(z_i)}\right\rangle_{\!\!\alpha,\beta}\\
	 = ~\delta^{d}\!\left(\sum_i p_i\right)~\sum_{\text{solns}} \int \frac{\diff^{d}\ell}{\ell^2} \,  J^{-1} \, \text{PT}(\alpha) ~ \text{PT}(\beta)~,
\end{multline}
when the external particles are in bi-colour structure $(\alpha,\beta)$. Here, the integral on the {\it lhs} is taken over the full moduli space, including both the moduli of  the $(n+2)$-punctured Riemann sphere and the moduli of the gauge field $e$.  On the {\it rhs}, the sum is taken over all non-singular solutions to the 1-loop scattering equations~\eqref{eqn:OneLoop-ScattEqns-Sphere}, weighted by the Jacobian 
\begin{equation}\label{eqn:Jacobian}
	J = \omega_{rst}^2 ~\det\left( \frac{\partial\, \text{SE}_i}{\partial z_j}  \right)
\end{equation}
where each $\omega_{rst}= S_{rs}S_{st}S_{tr}$ is the usual $\text{vol}\, \text{SL}(2;\mathbb{C})$ factor from the zero modes of $c$ and $\tilde c$, SE$_i$ is the $i^{\text{th}}$ scattering equation~\eqref{eqn:OneLoop-ScattEqns-Sphere-3} where $i,j\in\{1,\ldots,n+2\} / \{r,s,t\}$. This agrees with the expression for the integrand of the 1-loop amplitude in this bi-adjoint theory given in~\cite{Geyer:2015bja,Geyer:2015jch}.


 \section{The Yang-Mills gluing operator}\label{sec:YangMills}
 Having described in detail the use of the gluing operator in computing amplitudes in the bi-adjoint scalar theory, we now turn to Yang-Mills theory. In this section, we will concentrate on the case of 1-loop amplitudes.
 
 \medskip
 
 The main new ingredient in the ambitwistor string including Yang-Mills theory is (one copy of) the worldsheet fermion system $\psi$, providing the origin of the CHY Pfaffians. This has two main consequences. Firstly, the gluing operator must account for states in both the Neveu-Schwarz and Ramond sectors, allowing both gluons and gluinos to run around the loop.   Consequently, the gluing operator in Yang-Mills theory takes the form
\begin{equation}\label{eqn:GluingOperatorYM1}
 	\Delta_{\text{YM}}(z_+,z_-) \,=\, \int   \frac{\diff^d \ell}{\ell^2}  \,\left(\Delta_{\text{NS}}(z_+,z_-) +\Delta_{\text{R}}(z_+,z_-) \right) 
	c\,\tilde c\,j^a(z_+)\,\delta_{ab}\, c\,\tilde c\,j^b(z_-)\,
	W_{\text {YM}}(z_+,z_-) 
\end{equation}
where  $\Delta_{\text{NS}}$ and $\Delta_{\text{R}}$ are the contributions from the NS and R sectors, respectively. We will see that the worldsheet correlation function involving $n$ external gluon vertex operators as well as this gluing operator can be computed in closed form in both the NS and R sectors.

The second consequence is that there is a fermionic worldsheet symmetry which, in the absence of the gluing operator, is generated by the current $G = P\cdot\psi$. These currents have the OPE $G(z)\,G(w) \sim P^2/(z-w)$. Now, as we saw for the bi-adjoint scalar theory, the gluing operator depends on the bosonic gauge field $e$ and modifies the scattering equations. In the case of Yang-Mills, consistency with this fermionic symmetry thus demands that the gluing operator also depends on the fermionic gauge field $\chi$. The correct choice is perhaps not entirely obvious, and for $W_{\text {YM}}$ is given by
\begin{equation}\label{eqn:W-FactorTree-YM}
  W_{\text {YM}}(z_+,z_-)   = \exp \left(  \frac{\ell^2}{2} \int_\Sigma e(x) \, \omega_{+-}^2(x)+ \frac{\ell^2}{2}\int _{\Sigma \times \Sigma}  \chi(x) \,   \omega_{+-}(x)  \, S(x,y)  \, \chi(y) \,    \omega_{+-}(y)    \right) ~.
 \end{equation}  
This factor is common to both the NS and R sectors. (There will be further $\chi$ dependence in part of the gluing operator specific to the NS sectors.)

We remark that, just as in the bi-adjoint scalar theory, both $\Delta_{\text{NS}}$ and $\Delta_{\text{R}}$ depend on the field $X$ only through a factor $e^{i\ell\cdot(X(z_+)- X(z_-)}$. Thus, integrating out $X$ again leads to
\begin{equation}\label{eqn:P-onshell-oneloop-sphere-repeat}
   P(z) = \ell \, \omega_{+-}(z) + \sum_{i=1}^n p_i \, \omega_{i \ast}(z)  \, ,
 \end{equation}
and then integrating out the bosonic gauge field $e$ leads to the same 1-loop scattering equations
\begin{equation}\label{eqn:oneloop-scatteqns-repeat}
 P^2 = \ell^2 \, \omega_{+-}^2
 \end{equation}
as before. Again, these are the scattering equations that~\cite{Geyer:2015bja} required to describe 1-loop amplitudes in SYM theory.

Similarly, the only dependence that the YM gluing operator has on the current algebra associated to the target space gauge group is the common factor of $j^a(z_+)\,\delta_{ab}\,j^b(z_-)$. The current algebra correlator factors out of the correlation function and generates a sum over all possible Parke-Taylor factors in all orders, including multi-trace terms. As with the bi-adjoint scalar theory discussed above, we \emph{by hand} choose to extract only those single-trace terms in which $z_+$ and $z_-$ are adjacent in the colour ordering. Again, we expect that a fully consistent ambitwistor string for Einstein-Yang-Mills would possess a gluing operator that does not need such manipulation\footnote{It would also be interesting to investigate this in the context of the variant of ambitwistor string proposed in~\cite{Casali:2015vta} that describes a theory in which the Yang-Mills action is replaced by $\int {\rm tr}(d_Ab\wedge *F)$, where  $b$ is an adjoint-valued 1-form, independent of the connection $A$.}.


\subsection{Neveu-Schwarz sector}

We now describe the NS sector part of the gluing operator~\eqref{eqn:GluingOperatorYM1}. In this sector, 
\begin{equation}\label{eqn:GluingOperatorYM1NS}
    \Delta_{\text{NS}}(z_+,z_-) 
    =   \mathcal{O}^{\mu}(z_+) \,  \Delta_{\mu\nu}  \, \mathcal{O}^{\nu}(z_-)~,
 \end{equation}
where
\begin{subequations}\begin{equation}\label{eqn:GluingOperatorYMvector}
	\mathcal{O}^{\mu}(z_\pm) 
	= \delta(\gamma (z_\pm))  \left(\psi^\mu (z_\pm)  -\ell^\mu\int_\Sigma \chi(x) \,S(z_\pm,x)\,\omega_{+-}(x)\right) 
	e^{\pm i  \ell \cdot X(z_\pm)}
\end{equation}
describes the contribution of an off-shell vector mode, and the tensor structure is 
\begin{equation}
	\Delta_{\mu\nu}  =  \eta_{\mu\nu} - \frac{\ell_\mu\xi_\nu + \xi_\mu\ell_\nu}{\xi\cdot \ell} ~.
\end{equation}
\end{subequations} 
Here $\xi_\mu$ is an arbitrary vector that ensures only transverse modes propagate around the loop. One may check that the full gluing operator~\eqref{eqn:GluingOperatorYM1} (including the factor of $W(z_+,z_-)$) is BRST invariant under the transformations~\eqref{eqn:BRSTaction}\footnote{In fact, there is potential failure of BRST invariance in the NS sector arising on the boundary of the moduli space where $z_+=z_-$. We will see later that this failure is cancelled by the R sector.}. This BRST invariance in particular ensures that the amplitude is independent of the particular choice of $\xi$. 

\medskip

We now consider the path integral over the fermionic fields, which will yield a (reduced) CHY Pfaffian for $n+2$ particles. The gluing operator depends on $\chi$ through $\mathcal{O}^{\mu}$ as well as deforming the fermionic moduli through $W(z_+,z_-)$. Thus, as well as modifying the scattering equations as before (through its dependence on $e$), the insertion of this Yang-Mills gluing operator will also change the entries of the $A $ and $C$ block of the CHY Pfaffian.

 We choose to work in a picture where the $n$ external NS vertex operators are at picture zero, where the two zero modes of the $\gamma$ ghosts at genus zero are absorbed by the $\delta(\gamma)$ factors of the gluing operator. More explicitly, this statements means that we choose to expand the fermionic moduli in a basis
\begin{equation}
	\chi(z) ~= \sum_{r=1}^{(n+2)-2} \eta_r\, \chi_r(z)~,
\end{equation}
where for $r = 1,\ldots,n$, the $\eta_r$ are Grassmann valued constants, transforming as an element of $T^{1/2}_{z_r}$, and the moduli 
$$
	\text{span}( \{ \chi_r \}_{r=1}^{n} )= H^1(\Sigma,\,T^{1/2}_\Sigma(-z_1-\cdots -z_n - z_+-z_-))
$$ 
are chosen to extract the residue at the $r^{\rm th}$ marked point. That is, for any meromorphic  $f\in H^0(\Sigma,K^{3/2}_\Sigma(z_1+\cdots+z_n+z_++z_-))$, we have
\begin{equation}
	\int_\Sigma \chi_r(z) f(z) = \text{Res}_{r} f(z)~,
\end{equation}
where the residue is understood as an element of $K^{1/2}|_{z_r}$. (This pairing is an instance of Serre duality.) Note that there are exactly $n$ such moduli, two less than the total number of points marked by either the external vertex operators or the gluing operator. Thus, after performing the $\beta\gamma$ path integral, our choice of basis amounts to descending all $n$ external vertex operators to picture 0. In addition, since $S(z_\pm,x)\, \omega_{+-}(x)$ has no pole as $x$ approaches any of the external marked points (at least for generic choices of these points), we have
\begin{equation}
	\int_\Sigma \chi_r(x) \,S(z_\pm,x)\,\omega_{+-}(x) = 0 \qquad \forall \ r \,\in\,\{1,\ldots,n\}
\end{equation}
so with this choice of basis, both the shift of $\psi^\mu$ in~\eqref{eqn:GluingOperatorYMvector} and the fermionic contribution to $W(z_+,z_-)$ in~\eqref{eqn:W-FactorTree-YM} vanish. Consequently, in this basis the contribution from the $\beta\gamma$ and $\psi$ system becomes
\begin{equation}\label{eqn:NS-integrand}
 \Delta_{\mu\nu}  \left\langle   \delta(\gamma(z_+))\, \psi^\mu(z_+)   ~ \delta(\gamma(z_-))\,  \psi^\nu(z_-) ~ \prod_{i=1}^n (\varepsilon_i \!\cdot \!P(z_i) + \varepsilon_i \!\cdot\! \psi(z_i) \,  p_i\! \cdot\! \psi(z_i) ) \right\rangle_{\!\!\beta\gamma,\psi}
 \end{equation} 
where we recall that  $P(z_i)$ is frozen by~\eqref{eqn:OneLoop-FrozenP}. We remark that this correlator can be computed using any basis for the fermionic moduli -- so long as the full $\chi$ dependence of the gluing operator is taken into account, one finds the same result in any basis. In particular, this implies that the correlator is independent of the location of the PCOs, as for the tree amplitude. The curious reader may consult appendix \ref{apx:gluingCorrelators} for the evaluation of \eqref{eqn:NS-integrand} in an generic basis.

 \medskip

We now proceed to evaluate the correlator~\eqref{eqn:NS-integrand}. The $\beta\gamma$ path integral is trivially performed and gives a factor
\begin{equation}\label{eqn:BetaGamma}
	\left\langle \delta(\gamma(z_+)) ~ \delta(\gamma(z_-))\right\rangle_{\beta\gamma} = \frac{\sqrt{dz_+\,dz_-}}{z_+-z_-}=: S_{+-} ~.
\end{equation}
The more interesting part is the $\psi$ path integral. If not for the fact that the $\psi$ fields at $z_\pm$ are contracted into the target vector  propagator $\Delta_{\mu\nu}(\ell)$, this would readily evaluate to a simple Pfaffian. This motivates us to rewrite 
\begin{equation}
 \Delta_{\mu\nu}  ~\psi ^\mu(z_+)\, \psi ^\nu(z_-)  = \Delta_{\mu\nu} ~  \frac{\partial ^2}{\partial \epsilon^+_\mu ~ \partial \epsilon^-_\nu} ~~ \varepsilon^+ \! \cdot \! \psi(z_+)   ~ \varepsilon^- \! \cdot \! \psi(z_-)~,
 \end{equation}
 where $\varepsilon^\pm$ are auxiliary `polarization vectors' for the state flowing in the propagator -- note that, as usual, these off-shell polarization vectors are not required to be transverse to $\ell$. With this replacement, we can immediately evaluate the correlator \eqref{eqn:NS-integrand}. Including the $\beta\gamma$ contribution~\eqref{eqn:BetaGamma}, we find
\begin{equation}\label{eqn:NS-integrand-step2}
   S_{+-} \, \Delta_{\mu\nu}  ~  \frac{\partial ^2}{\partial \varepsilon^+_\mu ~ \partial \varepsilon^-_\nu}  ~~ \pf \left(   \begin{array}{cc|c}
0 & \varepsilon^+ \cdot \varepsilon^- ~ S_{+-} & \varepsilon^+ \cdot v_j ~ S_{+j}  \\ \cdot & 0 & \varepsilon^- \cdot v_j ~ S_{-j}  \\ \hline \cdot & \cdot & M
\end{array}   \right)  ~,
 \end{equation} 
 where $M$ is the full, $(2n)^{\times 2}$, tree level CHY matrix involving only the external particles, and $v_j$ represents either $p_j $ or $\varepsilon_j$ depending on whether the index $j$ is in the first of second half of $M$.  We also abbreviated $ S_{ij} \equiv S(z_i , z_j) $, and all indices are understood modulo $ n$ unless otherwise stated. To perform the derivatives, we use that the variation of a Pfaffian is given by
 \begin{equation}\label{eqn:PfaffianDerivative}
  \delta \, \pf ( M)  =   \sum_{i<j}\,  (-1)^{i+j+1} ~ \delta M_{ij}~ \pf \left(M ^{\check \imath  \check \jmath}\right) 
 \end{equation} 
 where $M ^{\check \imath  \check \jmath}$ denotes the matrix obtained by removing rows/columns $i,j$. We find that the correlator \eqref{eqn:NS-integrand} becomes
 \begin{equation}\label{eqn:NS-integrandintermediate}
 S_{+-}^2 \left( \tr(\Delta)   + \sum_{i< j}  v_i ^\mu \Delta_{\mu\nu}  \, v_j^\nu ~ \frac{S_{i+}S_{-j} - S_{i-}S_{+j} }{  S_{+-}}~\frac{\partial}{\partial M_{ij}}  \right)~ \pf(M) ~,
 \end{equation}
 where $  \Delta_{\mu}^\mu   = d-2 =8$ is the number of transverse polarization states in this NS sector.

 \medskip
 
To compare to the result of~\cite{Geyer:2015bja,Geyer:2015jch},  we must process it further. Using the $q$-expansion of the Szeg\'o kernel given in \cref{eqn:Szego-expansions} we find that, on the support of the 1-loop scattering equations~\eqref{eqn:OneLoop-ScattEqns-Sphere},
\begin{equation}
	\pf(M)=\pf(M_3)|_{q^0}
\end{equation}
so that the Pfaffian of the full, $2n \times 2n $ tree-level CHY matrix involving only the external states\footnote{Note that this Pfaffian does not vanish, since we are on the support of the 1-loop scattering equations $P^2(z) = \ell^2\omega_{+-}^2(z)$.} precisely agrees with the $\mathcal{O}(q^0)$ term in the expansion of the spin-structure 3 Pfaffian on the torus, expanded around the degeneration limit $q=0$. Thus, the first term of~\eqref{eqn:NS-integrandintermediate} can immediately be written in terms of an object appearing in the genus one result, localized to $q=0$. Furthermore, a straightforward (though somewhat tedious) calculation reveals that the contribution to \eqref{eqn:NS-integrandintermediate} coming from the part of $\Delta_{\mu\nu}$ that is proportional to the metric $\eta_{\mu\nu}$ yields
 \begin{equation}
\sum_{i<j}  \frac{S_{i+}S_{-j} - S_{i-}S_{+j} }{  S_{+-}}  ~ v_i\cdot v_{j} ~\frac{\partial}{\partial M_{ij}} ~ \pf(M)  \,=\,\pf (M_3)  |_{\sqrt q}   \,,
 \end{equation}
{\it i.e.}, the co-efficient of $\sqrt{q}$ in the $q$-expansion of the same Pfaffian as the torus degenerates.

Combining these two, our result~\eqref{eqn:NS-integrandintermediate} for the genus zero correlator including the NS sector part of the gluing operator can be written as
\begin{equation}\label{eqn:AnswerPlusJunk}
\begin{aligned}
	& S_{+-}^2\left(8 \, \pf(M_3)|_{q^0} + \pf(M_3)|_{\sqrt{q}}\right) \\
 &\hspace{1cm}- \, S_{+-}^2 \left[ \eta_{\mu\nu}   - \Delta_{\mu\nu}   \right] ~ \sum_{ i <  j}  v_i^\mu v_j^\nu ~ (S_{i+}S_{-j} - S_{i-}S_{+j} )~\frac{\partial}{\partial M_{ij}} ~ \pf(M)~.
\end{aligned}
\end{equation}
When multiplied by the appropriate current correlator (worldsheet Parke-Taylor factor), the first line here agrees precisely with the result of~\cite{Geyer:2015bja,Geyer:2015jch} for the NS sector contribution to the 1-loop integrand of SYM. We would thus like to show that the second line vanishes. 

Firstly, note that the tensor structure 
$$
	  \eta_{\mu\nu} -   \Delta_{\mu\nu}   = \frac{\ell_\mu\xi_\nu + \xi_\mu\ell_\nu}{\ell\cdot\xi}
$$ 
reveals that the second line in~\eqref{eqn:AnswerPlusJunk} comes from longitudinal states flowing through the propagator. To see that these states decouple, so that this line in fact vanishes, we rewrite it in the equivalent form
 \begin{equation}\label{eqn:unphys-state-decoupling-pfaffian}
  \frac{1}{  \ell\cdot\xi} \, S_{+-}^2 \,  \left[\pf \left(   \begin{array}{cc|c}
~0~ & ~0~~ & \ell \cdot v_j ~ S_{+j}  \\ \cdot & 0 & \xi \cdot v_j ~ S_{-j}  \\ \hline \cdot & \cdot & M
\end{array}   \right) + 
   \pf \left(   \begin{array}{cc|c}
~0~ & ~0~~ & \xi \cdot v_j ~ S_{+j}  \\ \cdot & 0 & \ell \cdot v_j ~ S_{-j}  \\ \hline \cdot & \cdot & M
\end{array}   \right)\right]  ~.
 \end{equation}
It is easily verified that each of these two Pfaffians vanishes on the support of the 1-loop scattering equations \eqref{eqn:OneLoop-ScattEqns-Sphere}, because e.g. the vector $V_j = (S^{-1}_{+-} , 0 | S^{-1}_{j-}  , 0_j )^{\rm T}$ is in the kernel of the first matrix, while $U_j = (0 , S^{-1}_{-+} | S^{-1}_{j+},0_j)^{\rm T}$ similarly lies in the kernel of the other. We emphasise that this \emph{does not} mean that the tensor structure in the gluing operator may simply be taken to be $\eta_{\mu\nu}$, since any choice of $\xi$ removes the two longitudinal modes. This can be seen explicitly in the factor of $8=d-2$ in the first line of~\eqref{eqn:AnswerPlusJunk}.

One can also see this decoupling of the two unphysical degrees of freedom directly at the level of the vertex operators. Let $\xi$ and $\xi'$ be two different choices of vector used to define the longitudinal part of the tensor structure of the propagator. Then
\begin{equation}\label{eqn:Different-Longitudes}
	\Delta_{\mu\nu}^{(\xi)} - \Delta_{\mu\nu}^{(\xi')} 
	= 2 \left(\frac{\ell_{(\mu}\xi'_{\nu)}}{\ell\cdot \xi'} -\frac{\ell_{(\mu}\xi_{\nu)}}{\ell\cdot \xi} \right)
\end{equation}
Thus, the difference between gluing operators defined with the help of $\xi$ and $\xi'$ is proportional to
\begin{equation}
	\delta(\gamma)\,\ell\cdot\psi_+ \,e^{i\ell\cdot X_+} ~~\delta(\gamma)\,\left[(\xi\cdot\ell) (\xi'\cdot\psi_-) - (\xi'\cdot\ell)(\xi\cdot\psi_-)\right]\,e^{-i\ell \cdot X_-} + (z_+\leftrightarrow z_-)~.
\end{equation}
The insertion at $z_+$ can be written as a BRST variation
 \begin{equation}
 \delta (\gamma) \, \, \ell  \!\cdot \! \psi ~ e^{i  \ell\cdot X }  = Q \circ \left( \delta'(\gamma) ~ e^{i  \ell \cdot X }    \right)  ~,
 \end{equation}
 where only the fermionic part of the BRST operator contributes non-trivially. However, in the difference of propagators with different longitudinal parts, the insertion at $z_-$ is actually BRST closed, since
 \begin{equation}
 \begin{aligned}
 &Q \circ \left( \delta(\gamma) \,  \left[(\xi\cdot\ell)(\xi' \cdot \psi)-(\xi'\cdot\ell)(\xi\cdot\psi)\right]   ~ e^{ -i  \ell\cdot X }  \,  \right) \\ 
 &\qquad\qquad= - \ \delta(\gamma)\,\partial\gamma\,\left[(\xi\cdot\ell)(\xi'\cdot\ell)-(\xi'\cdot\ell)(\xi\cdot\ell)\right]  \, e^{-i  \ell\cdot X }  = 0 ~,
 \end{aligned}
 \end{equation}	
 where the second line comes from double contractions, the factor of $\delta(\gamma)$ in the insertion annihilating the factor of $\gamma$ in the BRST operator at lowest order. Consequently, the difference between gluing operators with two different choices of longitudinal part is BRST exact, and decouples from any correlation function involving on-shell external states. Again, this does not mean that the tensor structure in the gluing operator may be taken to be $\eta_{\mu\nu}$.

 \medskip

To summarize, we have shown that the operator insertion \cref{eqn:GluingOperatorYM1NS} on the sphere gives rise to the NS sector contribution to the one-loop SYM integrand found by \cite{Geyer:2015bja,Geyer:2015jch}. In particular, we computed the CFT correlator with any number of gluon vertex operator insertions and demonstrated that the longitudinal degrees of freedom through the node decouple on the support of the scattering equations.


 \subsection{Ramond sector}
 As noted above, the $\psi$ system also has an associated Ramond sector that provides spin field vertex operators~\eqref{eqn:GluinoVertex} which describe gluinos. To account for the possibility that these gluinos run around the loop, the gluing operator also contains a contribution from the Ramond sector, to which we now turn. 
 
 \medskip

As a first attempt, one might assume that Ramond sector contribution to the SYM gluing operator should simply be
\begin{equation}\label{eqn:oneLoopOperator-R}
\Delta_{\text{R}}(z_+,z_-) \buildrel{?}\over{=} e^ {- \phi /2}  \, \Theta_\alpha (z_+) ~  e^{i  \ell \cdot X(z_+)}~( \tfrac{1}{2} \, C^{-1} \bar{\ell})^{\alpha \beta} ~e^ {- \phi /2}  \, \Theta_\beta(z_-) ~e^{-i  \ell \cdot X(z_-)}
 \end{equation}
which is a sum over the corresponding off-shell continuation $e^ {- \phi /2}  \,     \Theta_\alpha \, e^{i  \ell \cdot X}$ of the Ramond sector vertex operators inserted at $z_\pm$, joined by the tensor structure of the spin-$\sfrac{1}{2}$ propagator $ \tfrac{1}{2}\, (C^{-1}\bar\ell)^{\alpha\beta}$.

 However, while \eqref{eqn:oneLoopOperator-R} is correct, as it stands this operator cannot be added to the NS part of the propagator due to a mismatch in picture number -- the operator~\eqref{eqn:oneLoopOperator-R} has picture number $2 \times (-1/2 ) = -1$, while the NS part has $2 \times (-1 ) = -2$. Hence we change picture on one leg and write the Ramond sector contribution instead as 
 \begin{equation}\label{eqn:R-bispinor-insertion}
 	\Delta_{\text{R}}(z_+,z_-)  = 
	e^{-\phi/2}\,\Theta_\alpha (z_+)~  e^{i  \ell \cdot X(z_+)} ~(C^{-1} )\indices{^\alpha_{\dot \beta}}~ e^{-3\phi/2}\,\Theta^{\dot \beta}(z_-) ~  
	e^{-i  \ell \cdot X(z_-)}  ~,
 \end{equation}
 which has picture number $-1/2 -3/2 = -2$ and thus carries the right quantum numbers to be added to the NS contribution. One can check that both insertions \eqref{eqn:oneLoopOperator-R} and \eqref{eqn:R-bispinor-insertion} are indeed BRST closed, despite $\ell^2 \neq 0$, using the OPE's \eqref{eqn:SpinfieldOPEs} and \eqref{eqn:BRSTaction}.
 
 It is important to note however that the NS and R components of the gluing operator separately have a BRST anomaly arising on the boundary of the moduli space where $z_+ =  z_- $. To understand this, we consider the limit of the gluing operator as $z_+ \to z_-$: using the OPE of the fields involved we find the insertion
 \begin{equation}\label{eqn:GluingOperator-TadpoleOperator}
 \lim_{z_- \to z_+} \Delta_\text{YM}(z_+,z_-) = \int \frac{ \diff^{10} \ell}{\ell^2}~~ c \, \partial c \, \tilde{c } \, \partial \tilde{c} \, \delta(\gamma)\delta(\partial \gamma) \, (    8 - 8 )
 \end{equation}
 at the point where the two nodes meet, with the two contributions of opposite sign coming from the NS and R\footnote{This uses the bosonization identity $e^{-2 \phi} = \delta (\gamma) \delta (\partial\gamma)$.} sector respectively and we fixed $d=10$. In addition, the entire correlator comes multiplied by the cyclic sum over Parke-Taylor factors, whose limit,
 \begin{equation}
 \begin{aligned}
  &  \lim _{z_- \to z_+}  \, \sum_{i=1}^n \, \text{PT} ( \{ 1 , \cdots , i  , +, - , i+  1 ,\cdots , n\} ) \\
  & {} = - \frac{1}{2}\,  \sum_{i=1}^n \,  \omega_{i, i+1}(z_+) ~ \text{PT} ( \{ 1 , \cdots , i  , +, i +  1 ,\cdots , n\} ) ~  ,
 \end{aligned}
 \end{equation}
  is finite due to the $U(1)$ decoupling identity. Notice that that in this limit all dependence on the loop momentum $\ell$ drops out, apart from the overall prefactor.

 The operator insertion \eqref{eqn:GluingOperator-TadpoleOperator} might seem innocuous, but is actually dangerous, since it causes the bases that were used for the moduli of the gauge fields to become degenerate. Conversely, there are directions in the moduli space, of which the integrand is independent. This can be seen by considering that the integrand depends on the moduli only through the combinations
\begin{equation}
 \int _\Sigma e_0 \,  P^2 \qquad \text{and} \qquad \int _\Sigma  \chi_0 ~ \psi \cdot P  ~,
\end{equation}
 and noticing that $P^2$ and $\psi \cdot P$ are now finite at $z_+$. For instance, in this limit we have
 \begin{equation}
 \int _\Sigma  e_0 \, P^2 ~=~ \int_\Sigma e_0 \, \left( \sum_{i\neq j} p_i \cdot p_j \, \omega_{i\ast}\, \omega_{j\ast} \right) ,
 \end{equation}
 which is clearly finite at $z_+$ and has also become independent of $\ell$. To see that this pairing is now degenerate, we expand the field $e_0$ in a basis of some coordinates on the $n-1$ dimensional moduli space, e.g.
 \begin{equation}
 e (z) = \sum_{r=1}^{n-1} m^r \, e_r(z) \qquad \text{with} \qquad \int _\Sigma e_r \, P^2 = P^2(w_r)~.
 \end{equation}
  It is then easy to show that there are two linear combinations of the coordinates $m^r$ which drop out. The situation is completely analogous for the fermionic moduli. The integral over the moduli which the integrand is independent of produces a $0^2$ for the moduli of $\chi$, while those of $e$ contribute $\infty^2$. This can be regularized and the net contribution made finite, though potentially ambiguous.
 
 This anomaly is particularly problematic since $z_+ = z_-$ actually always arises in $(n-2)!$ of the solutions to the scattering equations, called `singular solutions'. One way to handle this is to discard certain singular solutions of the scattering equations, which effectively regularizes the operator insertion and is discussed in \cite{He:2015yua}. If however both NS and R parts are included, as in  \eqref{eqn:GluingOperator-TadpoleOperator}, the BRST ambiguity cancels between the sectors. This cancellation depends crucially on the relative coefficient between the NS and R term, and is a manifestation of the target space supersymmetry of the model.

\medskip

 We now compute the correlator of $n$ external NS sector vertex operators in picture 0, together with this R sector contribution to the gluing operator. This can be done using the remarkable, closed-form expression for the spin field correlator in the $\psi$ path integral obtained for example by Haertl, Schlotterer and Stieberger in \cite{Haertl:2009yf,Hartl:2010ks}. They find 
 \begin{equation}\label{eqn:RamondSectorCorrelatorRaw}
 \begin{aligned}
  & \qquad (C^{-1} )\indices{^\alpha_{\dot \beta}} ~  \bigg\langle \Theta_\alpha(z_+) \Theta^{\dot \beta } (z_-) \, \prod_{i=1}^{2n} v_i \cdot \psi (z_i) \bigg\rangle_{\psi}  \\
  {}={}&  S(z_+,z_-)^{5/4} \prod_{i=1}^{2n}\sqrt{\omega_{+-}(z_i)}  \sum_{m=0}^n 2^{-m} \!\!\!\! \sum_{\rho \in S_{2n}/ \mathcal{Q}_{n+1,m}} \text{sgn}(\rho) ~  \tr(  {v}_{\rho(1)}  {\bar{v}}_{\rho(2)}  \cdots \bar{v}_{\rho(2m)} )  \\
  & ~~~ \times \prod_{j=1}^{n-m} v_{\rho(2m+2j-1)} \cdot  v_{\rho(2m+2j)} ~ \frac{ S(z_{\rho(2m+2j-1)},z_{ \rho(2m+2j )}) \, S(z_+,z_-)}{S(z_{\rho(2m+2j-1)},z_+) \,  S(z_{\rho(2m+2j)},z_-) }~,
 \end{aligned}
 \end{equation}
 where the summation is taken over permutations 
\begin{equation}
 \begin{aligned}
  S_{2n}/ \mathcal{Q}_{n+1,m}  \equiv \{ \, & \rho \in S_{2n} : \, \rho(1) < \rho(2) < \cdots < \rho(2m) , \\
  & \rho(2m+2j-1) < \rho(2m+2j) ~ \forall j = 1,2, \cdots , n-m ~, \\
  & \rho(2m +2) < \rho(2m+4) < \cdots < \rho(2n) \, \}~.
 \end{aligned}
 \end{equation} 
 To avoid possible confusion we emphasize that in this formula there are $2n$ distinct marked points $z_i$ and vectors $v_i$, which we only later specialize to the particular configuration we have. Reference~\cite{Haertl:2009yf,Hartl:2010ks} actually treats the more general case of arbitrary genus, and where the spinor indices on $\Theta_\alpha$ and $\Theta^{\dot\beta}$ are left uncontracted. In our case these indices are joined using $C^{-1}$, which allows us to make considerable simplications in~\eqref{eqn:RamondSectorCorrelatorRaw}.

To begin, note that permutations in $S_{2n}/\mathcal{Q}_{n+1,m}$ which leave the first $2m$ indices fixed act trivially on $\tr(  {v}_{\rho(1)}  {\bar{v}}_{\rho(2)}  \cdots \bar{v}_{\rho(2m)} )$, and that the coefficient of this trace sums to give a Pfaffian. Thus the inner sum in~\eqref{eqn:RamondSectorCorrelatorRaw} becomes
  \begin{equation}\label{eqn:RamondCalculation-Step2}
 \begin{aligned}
  & \sum_{\rho \in S_{2n}/ \mathcal{Q}_{n+1,m}}  \text{sgn}(\rho) ~   \tr(  {v}_{\rho(1)}  {\bar{v}}_{\rho(2)}  \cdots \bar{v}_{\rho(2m)} )  \\
  &\qquad ~ \times \prod_{j=1}^{n-m} v_{\rho(2m+2j-1)} \cdot  v_{\rho(2m+2j)} ~ \frac{ S(z_{\rho(2m+2j-1)},z_{ \rho(2m+2j )}) \, S(z_+,z_-)}{S(z_{\rho(2m+2j-1)},z_+) \,  S(z_{\rho(2m+2j)},z_-) } \\
  {}&= \sum_{\substack{\mathfrak{b} \subset \{1, \cdots , 2n\} \\ |\mathfrak{b}  | = 2m} }  
  \tr\left( v_{i_1} \bar{v}_{i_2}\cdots \bar{v}_{i_{2m}} \right)_{i\in\mathfrak{b}}  ~ \pf \left( v_i \cdot v_j \,  \frac{ S_{ij} ~ S_{+-}}{S_{ \min(i,j) ,+} ~S_{\max(i,j),-}}    \right)_{i,j\in \mathfrak{b}^c}
 \end{aligned}
 \end{equation}
 where the sum in the final line runs over all ordered subsets $\mathfrak{b}$ of the complete set of indices $\{ 1 , \cdots , 2n\}$ of length $2m$. 
 
We can simplify much further still by writing
  \begin{equation}
   \tr ( {v}_1 \, \bar {v}_2 \, v_3  \cdots   \bar{v}_{2n})   = \tfrac{1}{2}  \tr ( \slashed{v}_1 \,  \slashed{v}_2  \cdots   \slashed{v}_{2n})  + \tfrac{1}{2}   \tr ( \slashed{v}_1 \,  \slashed{v}_2  \cdots   \slashed{v}_{2n} \, \Gamma_{d+1})  
 \end{equation}
and consider the contribution of the two terms separately. Due to the fact that the Clifford algebra is a representation of a fermionic QM system, we have the beautiful identities\footnote{The authors would like to thank Piotr Tourkine for bringing these identities to our attention. Incidentally, note the similarity of the matrix $V$ to the CHY matrices, given that $\text{sgn}(i- j)$ is the propagator in fermionic QM.} for the vector part
 \begin{subequations}\label{eqn:QM-Pfaffian}
 \begin{equation}
 \begin{gathered}
   \tr ( \slashed{v}_1    \cdots   \slashed{v}_{2n}) =   \tr(1)  ~  \pf ( V) ~, \\
    V_{ij} =  v_i \cdot v_j ~  \text{sgn}(i-j) ~, 
    \end{gathered}
 \end{equation}
and the axial part
 \begin{equation}
 \begin{gathered}
  \tr ( \slashed{v}_1  \cdots   \slashed{v}_{2n} \, \Gamma_{d+1})   =   \frac{ \tr(1)}{(d-1)!!} \, \int \diff^d \Psi_0 ~  \pf (  A) ~,  \\
  A_{ij} = v_i \cdot v_j ~  \text{sgn}(i-j) + v_i \cdot \Psi_0 \, v_j \cdot \Psi_0 ~,
   \end{gathered}
 \end{equation}
  \end{subequations}
of the trace of $2n$ gamma-matrices,  where $\tr (1) =  2^{d/2} = 32$. Here $\Psi_0^\mu$ is a set of $d$ Grassmann numbers. Both these identities may be derived by computing a fermionic QM path integral on the circle (or simply verified using the Clifford algebra, Berezin integration rules and recursive definition of the Pfaffian). 

The fact that both factors in~\eqref{eqn:RamondCalculation-Step2} can be written in terms of Pfaffians allows us to combine them. Recall that the Pfaffian of the sum of any two antisymmetric matrices $X,Y$ can be expressed as\footnote{This is most easily understood via the definition of the Pfaffian in terms of differential forms. Given a two-form $X$ on $\mathbb{R}^{2n}$, in coordinates $X = X_{ij} \, \diff x^i \wedge\diff x^j$, we have $\ast \pf(X) =  \tfrac{1}{n!}  X^{ n} $. Now, given a second two-form $Y$, we clearly have $$ \ast \pf(X+Y) =\tfrac{1}{n!}  (X+Y)^{ n} =    \tfrac{1}{n!}  X^n +  \tfrac{1}{(n-1)!}  X^{n-1} \wedge \tfrac{1}{1!} Y + \cdots +  \tfrac{1}{n!} Y^n ~, $$ thus generating the sum over all partitions into two ordered subsets in \cref{eqn:PfaffianSumExpansion}.}
 \begin{equation}\label{eqn:PfaffianSumExpansion}
 \pf (X+Y) ~ = \sum_{ \mathfrak{b} \in \, \substack{\text{ordered}\\ \text{subsets}}} \text{sgn}(\mathfrak{b} , \mathfrak{b}^c)~~ \pf ( X_{ij})_{ij \in \mathfrak{b} } ~~   \pf ( Y_{ij})_{ij \in \mathfrak{b}^c }    ~,
 \end{equation}
where the sum is over \emph{all} ways of splitting the indices on $X,Y$ into two ordered subsets $\mathfrak{b}$ and $\mathfrak{b}^c$, of any size. Using~\eqref{eqn:QM-Pfaffian}, this is exactly the form we have in the double sum in~\eqref{eqn:RamondSectorCorrelatorRaw}. Thus, including the prefactor of $\prod_{i=1}^{2n} \sqrt{\omega_{+-}(z_i)}\,$,  we can combine these sums into a  Pfaffian with entries
   \begin{multline}
    v_i\cdot v_j\,\sqrt{\omega_{+-}(z_i) \, \omega_{+-}(z_j)} ~ \left[ \frac{1}{2} ~ \text{sgn}(i-j) +  \, \frac{ \,S_{ij} ~ S_{+-}}{S_{ \min(i,j) ,+} ~S_{\max(i,j),-}}  \right]    \\ \equiv  v_i\cdot v_j\, S_{ij}~ \frac{1}{2}  \left( \sqrt{\frac{S_{i+}S_{j-}}{S_{i-}S_{j+}}}  + \sqrt{\frac{S_{j+}S_{i-}}{S_{j-}S_{i+}}}  \right)  ~,
 \end{multline}
for the vector part, and  a similar Pfaffian for the axial part. Altogether, we find that \cref{eqn:RamondSectorCorrelatorRaw} evaluates to
 \begin{equation}\label{eqn:Ramond-Combined-Pfaffian}
 \begin{aligned}
 & \frac{\tr(1)}{2} \, \pf \left[ v_i \cdot v_j \, \frac{ S_{ij}}{2}  \left( \sqrt{\frac{S_{i+}S_{j-}}{S_{i-}S_{j+}}}  + \sqrt{\frac{S_{j+}S_{i-}}{S_{j-}S_{i+}}}  \right)   \right]  \\
  &  +  \frac{\tr(1)}{2 \cdot 9!!} \int \diff ^d \Psi_0 ~ \pf \left[ v_i \cdot v_j \, \frac{S_{ij}}{2}  \left( \sqrt{\frac{S_{i+}S_{j-}}{S_{i-}S_{j+}}}  + \sqrt{\frac{S_{j+}S_{i-}}{S_{j-}S_{i+}}}  \right) + v_i \cdot \Psi(z_i) \, v_j \cdot \Psi(z_j)  \right]
 \end{aligned}
 \end{equation}
 where we defined $\Psi(z) = \sqrt{\omega_{+-}(z)} \, \Psi_0$. We recognize $\Psi(z)$ as the $q\to 0 $ limit of the fermionic zero mode of the field $\psi(z)$ on the torus in the odd spin structure. 
 
There is also a ghost contribution in the Ramond sector, which is fortunately much more straightforward. We have
 \begin{equation}
  \left\langle e^ {- \phi(z_+) /2}    ~ e^ {-  3\phi(z_-) /2}    \right\rangle_{\beta\gamma}  =   S_{+-}^{3/4}  ~,
  \end{equation} 
which combines with the factor of $S_{+-}^{5/4}$ from the spin field correlator. Notice in particular that the final answer is manifestly symmetric under exchange of $z_+$ and $z_-$, so that it does not matter which `end' of the gluing operator we write in picture $-3/2$.

\medskip

So far our calculation has been for generic insertions $v_i \cdot \psi(z_i)$ in \cref{eqn:RamondSectorCorrelatorRaw}. To compare to the 1-loop answer of~\cite{Geyer:2015bja,Geyer:2015jch} we specialize to the case of $n$ external  NS vertex operators of picture number 0. We find that the entries of the vector and axial Pfaffians in~\eqref{eqn:Ramond-Combined-Pfaffian} can be expressed in terms of the $\mathcal{O}(q^0)$ limits of the genus-one Pfaffians of~\cite{Adamo:2013tsa} in spin structures 2 and 1, respectively. Thus  we finally obtain the contribution to the one-loop CHY integrand in the Ramond sector:
\begin{equation}\label{eqn:integrandRamondContribution}
 - \frac{S_{+-}^2}{\ell^2} \left(   8\, \pf (M_2 )|_{q^0} \,+\,  \frac{8}{9!!}\,\pf(M_1)|_{q^0}   \right) ~.
 \end{equation} 
 The first term is in precise agreement with the results of \cite{Geyer:2015bja,Geyer:2015jch} for spin structure 2 part of the Ramond sector contribution. This spin structure is all that is needed for the amplitude where the external kinematics are chosen to lie in fewer than eight dimensions, which is the case considered by~\cite{Geyer:2015bja,Geyer:2015jch}. The final term in~\eqref{eqn:integrandRamondContribution} can easily be shown to follow from the $q\to0$ limit of the odd spin structure integrand in~\cite{Adamo:2013tsa}, and contributes when the external kinematics are in generic dimensions.
 
\medskip 

To summarize, we have computed the worldsheet correlator of our proposed gluing operator \eqref{eqn:R-bispinor-insertion} in both the NS and R sectors. In the single trace colour structure with colour order  $\alpha$, we obtain 
\begin{equation}\label{eqn:YMfullamplitude}
\begin{aligned}
\int &\left\langle \Delta_{\text{YM}}(z_+,z_-)~
	\prod_{i=1}^n \,c(z_i)\,\tilde c(z_i)~t_a^{(i)}j^a(z_i) ~\delta(\gamma(z_i))\,\varepsilon_i\!\cdot\! \psi(z_i)~e^{ip_i\cdot X(z_i)}\right\rangle_{\!\!\alpha}\\
	 &=~\delta^{10}\!\left(\sum_{i=1}^n p_i\right)\int \frac{\diff^{10}\ell}{\ell^2}\,\sum~ J^{-1}~\text{PT}(\alpha\,\cup\,\{z_+,z_-\})\left(\mathcal{I}_{\text{NS}}\,+\,\mathcal{I}_{\text{R}}\right)
\end{aligned}
\end{equation}
where the integral on the left is taken over the full moduli space of the theory -- including the moduli of the bosonic and fermionic gauge fields $e$ and $\chi$, as well as the locations of the punctures. On the right, $J$ is the Jacobian~\eqref{eqn:Jacobian} from solving the 1-loop scattering equations, PT is the Parke-Taylor factor evaluated on these solutions, and 
\begin{equation}\label{eqn:NSRcontributions}
\begin{aligned}
	\mathcal{I}_{\text{NS}} &=  S_{+-}^2\left[ 8\,\pf(M_3)|_{q^0}  \,+\, \pf(M_3)_{\sqrt{q}}\right]\\
	\mathcal{I}_{\text{R}} &= -S_{+-}^2\left[ 8\, \pf(M_2)|_{q^0} \,+\, \frac{8}{9!!} \, \pf(M_1)|_{q^0 }\right]
\end{aligned}
\end{equation}
Finally, the sum inside the integral is taken both over all solutions of the 1-loop scattering equations~\eqref{eqn:OneLoop-ScattEqns-Sphere} and also over the location of the insertion of the pair $\{z_+,z_-\}$ jointly into the worldsheet Parke-Taylor factor. This expression is in perfect agreement with the form of the 1-loop SYM integrand found in~\cite{Geyer:2015bja,Geyer:2015jch}, extended to be valid with generic external kinematics.


\section{The gluing operator for gravity}\label{sec:Gravity}

We turn finally to the gluing operator for the ambitwistor string description of gravity. Unlike the models of the previous section, which suffer from various anomalies and have unwanted extra states in their spectrum, the action~\eqref{eqn:GaugeFixedActionComplete} defines a consistent worldsheet theory describing Type II supergravity in $d=10$~\cite{Mason:2013sva,Adamo:2013tsa}. Thus, in this case, we will evaluate the full worldsheet correlator without needing to neglect any terms `by hand'.

Having seen the form of the 1-loop gluing operator in SYM, its form in supergravity is not difficult to guess. We have 
\begin{equation}\label{eqn:GravityGluing}
\begin{aligned}
	\Delta_{g}(z_+,z_-) 
	&= \int \frac{\diff^d \ell}{\ell^2}~  c(z_+)\tilde c(z_+) \,c(z_-)\tilde c(z_-)~W_\text{GR}(z_+,z_-)\\
	&\hspace{1cm}\times\,\left(\phantom{\tilde\Delta}\hspace{-0.3cm}\Delta_{\text{NS}}(z_+,z_-) + \Delta_{\text{R}}(z_+,z_-)\right)\!
	\left(\tilde\Delta_{\text{NS}}(z_+,z_-) + \tilde\Delta_{\text{R}}(z_+,z_-)\right) ~.
\end{aligned}
\end{equation}
Here, $\Delta_{\text{NS}}$ and $\Delta_{\text{R}}$ are  given by~\eqref{eqn:GluingOperatorYM1NS} and~\eqref{eqn:R-bispinor-insertion} respectively,  as in SYM. $\tilde\Delta_{\text{NS}}$ and $\tilde\Delta_{\text{R}}$ are given by exactly similar operators, but constructed from the tilded fermion system that in gravity replaces the currents $j^a$. Again, we remark that the chirality of the spinor $\tilde\zeta$ in $\tilde\Delta_{\text{R}}$ should be chosen opposite / the same as that of $\zeta$ in $\Delta_{\text{R}}$ to obtain the loop integrand in Type IIA / IIB supergravity. Finally, non-local part of the the gravitational gluing operator is
\begin{equation}\label{eqn:GravityW}
\begin{aligned}
	W_\text{GR}(z_+,z_-) = & \exp \left(  \frac{\ell^2}{2} \int_\Sigma e(x) \, \omega_{+-}^2(x)\right)\\
	\times~ & \exp\left(\frac{\ell^2}{2}\int _{\Sigma \times \Sigma}  \hspace{-0.2cm}\chi(x) \,   \omega_{+-}(x)  \, S(x,y)  \, \chi(y) \,    \omega_{+-}(y)    \right)  \\
	\times~ & \exp\left(\frac{\ell^2}{2}\int _{\Sigma \times \Sigma}   \tilde\chi(x)\,\omega_{+-}(x) \,S(x,y)\,\tilde\chi(y)\,\omega_{+-}(y)  \right) 
	\end{aligned}
\end{equation}
and depends on both the fermionic gauge fields $\chi$, $\tilde\chi$. As above, this reflects the fact that since the gluing operator modifies the bosonic scattering equations, so too must it modify both fermionic symmetries. 

As in SYM, individually the NS and R parts of the gluing operator fail to be BRST invariant, with the failure localised on the boundary of  moduli space where $z_+= z_-$, but this failure cancels between the two sectors. Again this can be understood as a consequence of target space supersymmetry, which is not manifest in our RNS description.

\medskip

The path integrals over the $\psi$ and $\tilde\psi$ systems (and the associated ghosts) can be performed independently, and give contributions identical to the corresponding terms in SYM. Altogether we obtain that the correlation function of the gravitational gluing operator and $n$ external NS sector particles (describing either gravitons, dilatons or the $B$-field) gives
\begin{equation}\label{eqn:Gravity1LoopIntegrand}
\begin{aligned}
	 \int &\left\langle \Delta_g(z_+,z_-)\,\prod_{i=1}^n c(z_i)\,\tilde c(z_i)\,\varepsilon_i\cdot\psi(z_i)\,\tilde\varepsilon_i\cdot\tilde\psi(z_i)\,e^{ip_i\cdot X(z_i)} \right\rangle \\
	&=~ \delta^{10}\!\left(\sum_{i=1}^n p_i\right) \int \frac{\diff^{10}\ell}{\ell^2}\, \sum\,J^{-1} 
	\left(\phantom{\tilde{\mathcal{I}}_{N}}\hspace{-0.5cm}\mathcal{I}_{\text{NS}}  
	+   \mathcal{I}_{\text{R}} \right)\, \left( \tilde{\mathcal{I}}_{\text{NS}}  +   \tilde{\mathcal{I}}_{\text{R}} \right)~.
\end{aligned}
\end{equation}
Here $J$ is again the Jacobian from solving the one-loop scattering equations, $\mathcal{I}_{\text{NS}}$ and $\mathcal{I}_{\text{R}}$ were given in~\eqref{eqn:NSRcontributions}, and $\tilde{\mathcal{I}}_{\text{NS}}$, $\tilde{\mathcal{I}}_{\text{R}}$ are exactly analogous but with the tilded polarization vectors. As always, the sum is over all solutions to the 1-loop scattering equations. In~\cite{Geyer:2015bja,Geyer:2015jch} it was shown that, this final expression is equivalent to both the 1-loop integrand of supergravity, and the genus 1 ambitwistor string calculation of~\cite{Adamo:2013tsa}. Here we have derived it working purely with correlation functions in a CFT at genus zero.


 \section{Conclusions}
 \label{sec:Conclusions}

In this paper, we have introduced the gluing operator for the ambitwistor string and demonstrated its use with several examples. In particular, we have shown how it gives rise to the building blocks that are used to build up a cubic scalar tree amplitude from lower points. Furthermore we have shown that the CHY loop integrands for Yang-Mills and gravity can be computed directly from a correlation function in the $g=0$ ambitwistor string on nodal Riemann surfaces. As well as vertex operators representing the external states, the correlation function involves the new operator $ \Delta ( z , w ) $, which plays the role of the propagator in the target space field theory. It is clear that the above procedure can be applied recursively, with more gluing operators connecting building blocks of lower genus and/or with less external punctures. The final stage would be a sum over all (compatible) graphs with the vertices given by a correlator with two gluing operator legs and one external puncture, at which point the scattering equations completely trivialize. Thus we interpolate between the purely combinatorial problem of enumerating Feynman graphs and the purely algebraic problem of finding solutions to the scattering equations. More importantly, we hope that the gluing operator generalizes to higher loops in that it provides a systematic approach, intrinsic to the worldsheet CFT and constrained by BRST invariance, to computing higher loop integrands from the sphere. Following this approach would exhibit the integrand as a manifestly rational function of the external kinematic data, which is a property that is non-trivial in the representation in terms of Riemann theta functions on higher genus curve.

 Finally, we note that the construction of higher point amplitudes by sewing together two Riemann surfaces, or higher loop amplitudes by self-sewing a Riemann surface, is exactly what one does in the operator approach to standard string theory (see, e.g.,~\cite{AlvarezGaume:1988bg,AlvarezGaume:1988sj}). Non-chiral strings contain an infinite number of states in their BRST spectrum, so the standard string propagator is rather difficult to handle. Since the ambitwistor string spectrum is just that of a massless field theory, its propagator should be correspondingly simpler.  Indeed, it seems certain that the gluing operator we have presented here is nothing other than the string propagator, computed according to the general method  of~\cite{Cohen:1985sm,Birmingham:1987ny}, but specialised to the case of the ambitwistor string worldsheet action. Thus, perhaps the most pressing question arising from this work is to find an {\it ab initio} derivation of the gluing operator as the ambitwistor string propagator. Part of the attraction of the operator approach to superstrings was that it allowed one easier access to higher genus worldsheet correlators. Hopefully, understanding the ambitwistor string propagator should similarly allow one to use the gluing operator to construct multi-loop integrands. We note that the operator approach to ambitwistor strings has begun to be explored in~\cite{Reid-Edwards:2015stz,Reid-Edwards:2017goq}; it would be interesting to relate that work to the ideas of this paper.

\vspace{1cm}

\noindent {\large{\bf Acknowledgments}}
 
\noindent It is a pleasure to thank Tim Adamo, Lionel Mason, Piotr Tourkine, Ricardo Monteiro and Yvonne Geyer for helpful conversations. The work of KR and DS is supported in part by a Marie Curie Career Integration Grant (FP/2007-2013/631289). The work of DS is supported in part by STFC consolidated grant ST/L000385/1.

 \pagebreak
 \appendix

\section{Sphere Correlator with NS Gluing Operator Insertion}\label{apx:gluingCorrelators}
The insertion of the gluing operator with $n$ fixed vertex operators for external particles requires the insertion of $n$ picture changing operators, at arbitrary (not necessarily distinct) locations $x_r$ on the sphere, as explained e.g. in \cite{Witten:2012bh}. As mentioned in the above, the choice of these $n$ arbitrary locations amounts to a choice of basis on the tangent bundle to the fermionic moduli space. Clearly the correlator should be independent of our choice of basis, and thus independent of the locations $x_r$ at which we insert the picture changing operators.

 In a generic basis, the correlator is given by
\begin{equation}\label{eqn:fermionCorrelatorGluingOperatorGenericBasis}
\begin{aligned}
 \Delta_{\mu\nu}~ &  \left\langle  ~~ \delta(\gamma(z_+))  \, \left( \psi^\mu(z_+) - \ell^\mu \int \chi_0(w_+) \, \omega_{+-}(w_+) \, S(w_+, z_+) \right)   \right. \\
    &  \qquad \delta(\gamma(z_-))  \, \left( \psi^\nu(z_-) - \ell^\nu \int \chi_0(w_-) \, \omega_{+-}(w_-) \, S(w_-, z_-) \right)      \\ 
    &  \qquad \exp \left(   \frac{\ell^2}{2}\int _{\Sigma \times \Sigma}  \chi_0(y) \,   \omega_{+-}(y)  \, S(y,y^\prime)  \, \chi_0(y ^\prime) \,    \omega_{+-}(y^\prime)    \right)  \\
 & \qquad \left.     ~ \prod_{i=1}^n \delta(\gamma(z_i)) \, \varepsilon_i  \cdot\! \psi(z_i) \qquad    \prod_{r=1}^n \, \delta(\beta(x_r)) \, P(x_r) \cdot \psi (x_r) ~~ \right\rangle_{\!\!\beta\gamma,\psi,\chi_0} ~,
\end{aligned}
 \end{equation} 
 where the first three lines contain the relevant parts of the gluing operator and the final line contains the external vertex operators as well as the picture changing operators. Recall that the field $P(z)$ is frozen to its classical value \cref{eqn:P-onshell-oneloop-sphere-repeat}. Notice that the expression \eqref{eqn:NS-integrand} is the limit of \eqref{eqn:fermionCorrelatorGluingOperatorGenericBasis} where one of each picture changing operators is taken to coincide with one of each of the vertex operators of the external particles. Without taking this limit, keeping the picture changing operators at generic positions, the correlator \eqref{eqn:fermionCorrelatorGluingOperatorGenericBasis} evaluates to
 \begin{equation}\label{eqn:evaluatedFermionCorrelatorGluingOperatorGenericBasis}
   \frac{V \big ( \{ z_i \} _{i=1}^{n,+,-}  \, | \,  \{ x_r \} _{r=1}^n \big) }{V \big(  \{ z_i \}_{i=1}^{n,+,-}  \big) ~ V \big( \{ x_r \} _{r=1}^n  \big)  } ~~ \Delta_{\mu\nu}  ~  \frac{\partial ^2}{\partial \varepsilon^+_\mu ~ \partial \varepsilon^-_\nu}  ~~ \pf \left(   \begin{array}{cc}
A  & - C^T \\   C & B
\end{array}   \right)  ~,
 \end{equation} 
with the familiar Vandermonde factors
\begin{equation}
\begin{aligned}
V \big(  \{ z_i \}_{i=1}^{n,+,-}  \big)  =  & \prod_{i < j =1}^{n,+,-} \, \frac{1}{S(z_i , z_j)}  ~, \qquad V \big(  \{ x_r \}_{r=1}^n  \big) = \prod_{r < s =1}^n \, \frac{1}{S(x_r , x_s)}  ~, \\
 &   V \big ( \{ z_i \} _{i=1}^{n,+,-}   \, | \,   \{ x_r \} _{r=1}^n \big) = \prod_{i=1}^{n,+,-} \prod_{r=1}^n \frac{1}{S(z_i , x_r)}  ~,
\end{aligned}
\end{equation}
and the matrix entries
\begin{equation}
\begin{aligned}
A_{rs} & = \left( P ( x_r) \cdot P(x_s) - \tfrac{\ell^2}{2}  \, \omega_{+-} ( x_r)\, \omega_{+-} ( x_s) \right) \, S(x_r , x_s)  ~, ~~ &  \\
B_{ij}  = \varepsilon_i & \cdot   \varepsilon_j  ~ S(z_i , z_j)  ~, \qquad ~~ C_{ir}  = \varepsilon_i \cdot \left( P(x_r) - \ell \, \omega_{+-} (x_r) \right) ~ S(z_i , x_r ) ~,
\end{aligned}
\end{equation}
for $ r,s = 1, \cdots , n $ and $ i,j = 1, \cdots , n , + , - $. It is readily shown that   \eqref{eqn:evaluatedFermionCorrelatorGluingOperatorGenericBasis} is independent of the locations $x_r$, by checking that it is free of poles as any $x_r \to x_s $ as well as $x_r \to z_i $, on the support of the scattering equations \eqref{eqn:oneloop-scatteqns-repeat}, and appealing to Liouville's theorem. Again, notice that the expression \eqref{eqn:NS-integrand-step2} is simply the limit of \eqref{eqn:evaluatedFermionCorrelatorGluingOperatorGenericBasis} where one of each of the $x_r$ is taken to coincide with one of each of the $z_i$.

\section{Degeneration limit of the torus Szeg\'o kernel}
The genus-one Szeg\'o kernels have following expansions at small $q = exp (i \pi \tau)$:
\begin{align}\label{eqn:Szego-expansions}
S_1 (z_i,z_j|\tau) &  = S_{ij} ~ \frac{1}{2} \left( \sqrt{\frac{S_{i+}S_{j-}}{S_{i-}S_{j+}}}  + \sqrt{\frac{S_{j+}S_{i-}}{S_{j-}S_{i+}}} \,  \right)  + \order {\sqrt q} {}  \\
S_2 (z_i,z_j|\tau) &  =  S_{ij} ~ \frac{1}{2} \left( \sqrt{\frac{S_{i+}S_{j-}}{S_{i-}S_{j+}}}  + \sqrt{\frac{S_{j+}S_{i-}}{S_{j-}S_{i+}}}  \,  \right)  + \order {\sqrt q} {}  \\
S_3 (z_i,z_j|\tau) & = S_{ij} + \sqrt{q} ~ \frac{S_{i+}S_{-j} - S_{i-}S_{+j} }{ S_{+-}} + \order q {}   \\
S_4 (z_i,z_j|\tau) &  = S_{ij} - \sqrt{q} ~ \frac{S_{i+}S_{-j} - S_{i-}S_{+j} }{ S_{+-}}+ \order q {}
\end{align}
where
\begin{equation}
S_{ij} \equiv S(z_i, z_j) = \frac{\sqrt{ \diff z_i \, \diff z_j } }{z_i - z_j}
\end{equation}
is the Szeg\'o kernel on the sphere and $z_\pm$ are the coordinates of the node. We can take into account the fermionic zero mode in spin structure $1$ by modifying the propagator of fermions, e.g. at $q=0$ (on the nodal sphere) we find
\begin{equation}
\eta^{\mu\nu} \, S_{ij} ~ \frac{1}{2} \left( \sqrt{\frac{S_{i+}S_{j-}}{S_{i-}S_{j+}}}  + \sqrt{\frac{S_{j+}S_{i-}}{S_{j-}S_{i+}}} \, \right)  + \sqrt{\omega_{+-}(z_i)\, \omega_{+-}(z_j)} ~  \psi_0^\mu \, \psi_0^\nu \\
\end{equation}
where $\psi_0^\mu$ is a set of $d$ Grassmann numbers, over which the entire correlator is to be integrated using the standard Berezin integration rules. Here $\omega_{+-}(z)$ is the $q\to 0$ limit of the unique holomorphic one-form on the torus, normalized to have residues $\pm 1$ at $z = z_\pm$, i.e.
\begin{equation}
\omega_{+-} (z)= \frac{\diff z ~ (z_+-z_-)}{(z-z_+)(z-z_-)}
\end{equation}
More details on the treatment of the zero mode can be found in \cite{Adamo:2013tsa} and many other places.


\bibliography{main}
\bibliographystyle{JHEP}

\end{document}